\documentclass[amsmath,amssymb,showkeys,superscriptaddress,floatfix,aps,prd,10pt]{revtex4}
\usepackage{graphicx,epstopdf}
\usepackage[caption=false]{subfig}
\usepackage{multirow}
\usepackage{appendix}
\usepackage{xcolor}
\usepackage[colorlinks=true, urlcolor=blue, linkcolor=blue, citecolor=green]{hyperref}
\def\pslash{p\!\!\!\slash }
\def\qslash{q\!\!\!\slash }
\def\xslash{x\!\!\!\slash }

\begin{document}

\title{Gravitational form factors of  hyperons in light-cone QCD}
%

\author{
 U. \"{O}zdem$^{1, 3, *}$
     and    
K. Azizi$^{2, 3, \dag}$ 
\\
  \small$^1$ Health Services Vocational School of Higher Education, Istanbul Aydin University, Sefakoy-Kucukcekmece, 34295 Istanbul, Turkey\\
 \small $^2$ Department of Physics, University of Tehran, North Karegar Avenue, Tehran 14395-547, Iran\\
\small $^3$ Department of Physics, Dogus University, Acibadem-Kadikoy, 34722 Istanbul, Turkey\\
 $^*$ e-mail:ulasozdem@aydin.edu.tr\\
 $^{\dag}$ e-mail:kazem.azizi@ut.ac.ir
 }​

\begin{abstract}
The quark parts of the gravitational form factors of hyperons are calculated by means of the light-cone QCD sum rule. In the calculations, the distribution amplitudes (DAs) of $\Sigma$, $\Xi$ and  $\Lambda$ together with the general forms of their interpolating currents as well as the quark part of the energy-momentum tensor current are used. These form factors can provide information on their mass and distributions of the angular momentum, energy and pressure inside the hyperons. It is obtained that the $Q^2$ dependencies of the hyperon  gravitational form factors are nicely characterized by a multipole fit function. Using the fits of these form factors, some mechanical properties such as the mechanical radius of the hyperons and the pressure and energy distributions at the center of these particles are obtained.
The obtained results can help us in better understanding of the internal structures of these baryons and the QCD as theory of the strong interaction.
\end{abstract}
 \date{\today}
\maketitle

\section{Motivation} 
The gravitational form factors of baryons are fundamental objects that provide valuable information on different observables related to the structure and mechanical properties of these particles. They are described by the help of the matrix element of the energy-momentum tensor \cite{Pagels:1966zza}. The matrix element of the quark part of the energy-momentum tensor current between two hyperonic states is defined as \cite{Polyakov:2018zvc},
\begin{widetext}

\begin{align}\label{matFFs}
  \langle H(p^\prime,s')|T_{\mu\nu}^q|H(p,s)\rangle &=
 \bar{u}_{H}(p^\prime,s')\bigg[A^{H-H}(Q^2)\frac{ \tilde P_\mu \tilde P_\nu}{m_{H}}
 +~i J^{H-H}(Q^2)\frac{(\tilde P_\mu \sigma_{\nu\rho}+\tilde P_\nu \sigma_{\mu\rho})\Delta^\rho}{2\,m_{H}}
    \nonumber\\
  & +~D^{H-H}(Q^2) \frac{\Delta_\mu \Delta_\nu- g_{\mu\nu} \Delta^2}{4\,m_{H}}
  +~ \bar c^{H-H} (Q^2) m_{H} g_{\mu \nu} \bigg]  u_{H}(p,s),
 \end{align}

\end{widetext}
for $\Sigma-\Sigma$, $\Xi-\Xi$ and $\Lambda-\Lambda$ transitions. The matrix element for $\Sigma^0-\Lambda$ transition can be obtained by the replacement $m_H \rightarrow (m_{\Sigma^0}+m_{\Lambda})/2$. In the above equation, $ A^{H-H}(Q^2) $, $ J^{H-H}(Q^2) $, $ D^{H-H}(Q^2) $ and $  \bar c^{H-H} (Q^2)$ are gravitational form factors, $\tilde P= (p'+p)/2$,   $\Delta = p'-p$,  $q = p-p'$,  
$\sigma_{\mu\nu}= \frac{i}{2}[\gamma_\mu, \gamma_\nu]$ and  $Q^2=- \Delta^2$.
By calculation of the gravitational form factors one can calculate the pressure, shear forces, 
total angular momentum and spatial distribution of energy inside the particles \cite{Polyakov:2002yz}.
Calculation of gravitational form factors of hadrons is relatively new subject although they were introduced  by Kobzarev and Okun
 in 1962 \cite{Kobzarev:1962wt}. 
  A number of contributions have been made to the calculation of these form factors, recently.
Thus, the gravitational form factors of the nucleon have been extracted within various theoretical approaches such as, chiral perturbation theory ($\chi$PT) \cite{chen:2001pva, Belitsky:2002jp, Ando:2006sk, Diehl:2006ya, Diehl:2006js, Dorati:2007bk},  Bag model~\cite{Neubelt:2019sou}, instanton picture (IP)~\cite{Polyakov:2018exb},  chiral quark soliton model ($\chi$QSM) \cite{Petrov:1998kf, Schweitzer:2002nm, Ossmann:2004bp, Wakamatsu:2005vk, Wakamatsu:2006dy, Goeke:2007fq,Goeke:2007fp, Jung:2013bya,Jung:2014jja, Jung:2015piw,Wakamatsu:2007uc}, dispersion relation (DR)~\cite{Pasquini:2014vua}, 
Skyrme model \cite{Cebulla:2007ei,Kim:2012ts}, lattice QCD \cite{Hagler:2003jd,mathur:1999uf,  Gockeler:2003jfa, Bratt:2010jn, Hagler:2007xi,Brommel:2007sb,Negele:2004iu,Deka:2013zha},  light-cone QCD sum rules (LCSR)~\cite{Anikin:2019kwi,Azizi:2019ytx}  and   instant-front form (IFF) \cite{Lorce:2018egm}.
Recently, as a first calculation,  the transition gravitational form factors between $ N^* $ and $ N $ baryons  have been calculated in the framework of  light-cone QCD sum rule \cite{Ozdem:2019pkg}.

The hyperon states are very interesting as they can be used to check the order of SU(3) flavor symmetry breaking at different interactions. 
Our knowledge on these particles is very limited because of unstable nature of these particles. This makes experimental studies very difficult and this situation increase the importance of theoretical studies on different aspects of hyperons. In this accordance,
 we investigate the gravitational form factors defining the $\Sigma-\Sigma$, $\Xi-\Xi$, $\Lambda-\Lambda$ and the $\Sigma^0-\Lambda$ transitions within the light-cone QCD sum rule approach \cite{Braun:1988qv, Balitsky:1989ry, Chernyak:1990ag}.
To our best knowledge, this is the first study on the gravitational form factors of the light octet hyperons  in  the literature.
However, there have been attempts to extract the octet and decuplet hyperon's electromagnetic form factors \cite{Carrillo-Serrano:2016igi, Liu:2013fda, Jiang:2009jn, Lin:2008mr,  Ramalho:2011pp, Aliev:2013jta, Aliev:2002ra, Aliev:2009pd, Aliev:2013jda, Aliev:2013mda, Aliev:2000cy}, axial form factors \cite{Lin:2008mr, Lin:2007ap, Ramalho:2015jem, Ledwig:2014rfa, Kucukarslan:2014mfa,  Erkol:2011qh, Choi:2010ty, Erkol:2009ev, Jiang:2009sf, Jiang:2008aqa} as well as their tensor form factors \cite{Ledwig:2010tu, kucukarslan:2016xhx}.

The outline of the paper is as follows: In Sec. \ref{secII}, using LCSR, analytical results are obtained for hyperon's gravitational form factors.  Sec. \ref{secIII} includes the numerical analyses and 
Sec. \ref{secIV} is devoted  to summary and concluding remarks.

 \section{Hyperon's Gravitational Form Factors}\label{secII}

To derive the light-cone QCD sum rules for gravitational form factors of hyperons, we introduce the correlation function
\begin{align}\label{corf}
\Pi_{\mu\nu}(p,q)=i\int d^4 x e^{iqx} \langle 0 |\mathcal{T}[J_{H}(0)T_{\mu \nu}^q(x)]|H(p)\rangle,
\end{align}
where $T_{\mu\nu}^q(x)$ is the quark part of the symmetric energy-momentum tensor current and 
$J_H(0) $ are interpolating currents for hyperon states. The explicit forms of the $J_H(0) $ for different members and $T_{\mu\nu}^q(x)$ are given as
%
    %

 \begin{eqnarray}\label{hypcurr}
	J_\Sigma(0)&=&2\epsilon^{abc}\sum_{\ell=1}^{2}(u^{aT}(0) C J_1^\ell s^b(0))J_2^\ell u^c(0),\nonumber\\
	J_\Xi(0) &=& J_\Sigma(0) (u\leftrightarrow s), \nonumber\\
	J_\Lambda(0)&=&\frac{2}{\sqrt 6}\epsilon^{abc}\sum_{\ell=1}^{2}\Big[2(u^{aT}(0) C J_1^\ell d^b(0))J_2^
\ell s^c(0)+ (u^{aT}(0) C J_1^\ell s^b(0))J_2^\ell d^c(0)- (d^{aT}(0) C J_1^\ell s^b(0))J_2^\ell u^c(0)\Big],\nonumber\\
J_{\Sigma^0}(0)&=&\sqrt{2}\epsilon^{abc}\sum_{\ell=1}^{2}\Big[(u^{aT}(0) C J_1^\ell s^b(0))J_2^\ell d^c(0)
+(d^{aT}(0) C J_1^\ell s^b(0))J_2^\ell u^c(0)\Big],
\end{eqnarray}
 and 
 \begin{eqnarray}\label{tenscur}
  T_{\mu\nu}^q (x) &=& \frac{i}{2}\bigg[\bar{u}^d(x)\overleftrightarrow{D}_\mu(x) \gamma_\nu u^d(x) 
 + \bar{d}^e(x)\overleftrightarrow{D}_\mu(x) \gamma_\nu d^e(x) 
+(\mu \leftrightarrow \nu) \bigg],
\end{eqnarray}
where $J_1^1=I$, $J_1^2=J_2^1=\gamma_5$,  $J_2^2=t$, $\overleftrightarrow{D}_\mu(x)$ is the two-sided covariant derivative, C is charge conjugation operator,  $t$ is arbitrary mixing parameter; an a, b, c, d, and e are color indexes. In the above equations, by $ \Sigma $ we mean $ \Sigma^+ $.

To obtain the gravitational form factors in light-cone QCD sum rule, we need to calculate the correlation function in two different frameworks. First, we calculate the correlation function in terms of hadronic parameters, known as hadronic side.  Second, we obtain the correlation function in terms of the quark-gluon parameters via distribution amplitudes of the related hadron, known as QCD side.
In order to suppress undesired contributions, which are related to  the excited and continuum states, the Borel transformations are applied.
 Then, we match both representations of the correlation function by the help of the quark-hadron duality ansatz.

To calculate the correlation function in terms of hadronic properties, a complete hadronic set with the same quantum numbers as the interpolation currents is inserted. Thus, the correlation function takes the form

 \begin{align}\label{phys}
 \Pi_{\mu\nu}^{Had}(p,q) &= \frac{\langle0|J_H(0)|{H(p',s')}\rangle\langle {H(p',s')}
 |T_{\mu \nu}^q(x)|H(p,s)\rangle}{m^2_{H}-p'^2}
 +...,
\end{align}
where 
\begin{align}\label{res}
\langle0|J_H(0)|{H(p',s')}\rangle = \lambda_H u_{H}(p',s'),
\end{align}
with $ \lambda_H $ and $ u_{H}(p',s') $ being the residue and Dirac spinor corresponding to the hyperon $ H $, respectively.
Substituting Eq. (\ref{matFFs}) and Eq. (\ref{res}) into Eq. (\ref{phys}), we acquire the
hadronic representation of the correlation function in terms of the hadronic observables as

\begin{widetext}
\begin{align}\label{hads}
  \Pi_{\mu\nu}^{Had}(p,q)  &= \frac{ \lambda_H}{m_H^2-p'^{2}}(\pslash^{\prime}+m_H)
\bigg[A^{H-H}(Q^2)\frac{ \tilde P_\mu \tilde P_\nu}{m_{H}}
 +~i J^{H-H}(Q^2)\frac{(\tilde P_\mu \sigma_{\nu\rho}+\tilde P_\nu \sigma_{\mu\rho})\Delta^\rho}{2\,m_{H}}
    \nonumber\\
  & +~D^{H-H}(Q^2) \frac{\Delta_\mu \Delta_\nu- g_{\mu\nu} \Delta^2}{4\,m_{H}}
  +~ \bar c^{H-H} (Q^2) m_{H} g_{\mu \nu} \bigg]  u_{H}(p,s).
 \end{align}
\end{widetext}


The next step is computation of the correlation function in terms of the QCD degrees of freedom as well as the hyperon's distribution amplitudes.
%
 Inserting the explicit expressions for the hyperon interpolating currents as well as the quark part of the energy-momentum current into
 Eq. (\ref{hypcurr}), and performing all contractions among the quark fields with the help of Wick theorem, for the correlation function in x-space, we obtain 
 \begin{widetext}
 
\begin{align}\label{QCD1}
	\Pi_{\mu\nu}^{QCD}(p,q)&=-\int d^4 x e^{iqx}\Bigg[\bigg\{ (\gamma_5)_{\gamma\delta}\, C_{\alpha\beta}\, ( \overleftrightarrow{D}_\mu (x)\gamma_\nu)_{\omega \rho} 
	+
	t\, (I)_{\gamma\delta}\,(C \gamma_5)_{\alpha\beta}\,( \overleftrightarrow{D}_\mu(x) \gamma_\nu)_{\omega \rho}\nonumber\\
	&+
	(\gamma_5)_{\gamma\delta} \, C_{\alpha\beta} \, ( \overleftrightarrow{D}_\nu (x)\gamma_\mu)_{\omega \rho} 
    +
	t\,(I)_{\gamma\delta}\,(C \gamma_5)_{\alpha\beta}\, ( \overleftrightarrow{D}_\nu (x)\gamma_\mu)_{\omega \rho}
	\bigg\}\nonumber\\
   & \times	
  \Big (\delta_\sigma^\alpha \delta_\theta^\rho \delta_\phi^\beta S_u(-x)_{\delta \omega}
     +\, \delta_\sigma^\delta \delta_\theta^\rho \delta_\phi^\beta S_u(-x)_{\alpha \omega}\Big)
     \langle 0|\epsilon^{abc} u_{\sigma}^a(0) u_{\theta}^b(x) s_{\phi}^c(0)|\Sigma(p)\rangle 
   \Bigg],
\end{align}

\end{widetext}
 for $\Sigma$-$\Sigma$ transition,

 \begin{widetext}
 
\begin{align}\label{QCD2}
	\Pi_{\mu\nu}^{QCD}(p,q)&=-\int d^4 x e^{iqx}\Bigg[\bigg\{ (\gamma_5)_{\gamma\delta}\, C_{\alpha\beta}\, ( \overleftrightarrow{D}_\mu (x)\gamma_\nu)_{\omega \rho} 
	+
	t\, (I)_{\gamma\delta}\,(C \gamma_5)_{\alpha\beta}\,( \overleftrightarrow{D}_\mu(x) \gamma_\nu)_{\omega \rho}\nonumber\\
	&+
	(\gamma_5)_{\gamma\delta} \, C_{\alpha\beta} \, ( \overleftrightarrow{D}_\nu (x)\gamma_\mu)_{\omega \rho} 
    +
	t\,(I)_{\gamma\delta}\,(C \gamma_5)_{\alpha\beta}\, ( \overleftrightarrow{D}_\nu (x)\gamma_\mu)_{\omega \rho}
	\bigg\}\nonumber\\
   & \times	
     \delta_\sigma^\alpha \delta_\theta^\delta \delta_\phi^\rho S_u(-x)_{\beta \omega} 
    \,\langle 0|\epsilon^{abc} s_{\sigma}^a(0) s_{\theta}^b(0) u_{\phi}^c(x)|\Xi(p)\rangle
   \Bigg],
\end{align}

\end{widetext}
 for $\Xi$-$\Xi$ transition,

 \begin{widetext}
 
\begin{align}\label{QCD3}
	\Pi_{\mu\nu}^{QCD}(p,q)&=-\frac{1}{\sqrt{6}}\int d^4 x e^{iqx}\Bigg[\bigg\{ (\gamma_5)_{\gamma\delta}\, C_{\alpha\beta}\, ( \overleftrightarrow{D}_\mu (x)\gamma_\nu)_{\omega \rho} 
	+
	t\, (I)_{\gamma\delta}\,(C \gamma_5)_{\alpha\beta}\,( \overleftrightarrow{D}_\mu(x) \gamma_\nu)_{\omega \rho}\nonumber\\
	&+
	(\gamma_5)_{\gamma\delta} \, C_{\alpha\beta} \, ( \overleftrightarrow{D}_\nu (x)\gamma_\mu)_{\omega \rho} 
    +
	t\,(I)_{\gamma\delta}\,(C \gamma_5)_{\alpha\beta}\, ( \overleftrightarrow{D}_\nu (x)\gamma_\mu)_{\omega \rho}
	\bigg\}\nonumber\\
   & \times	\bigg\{
  \Big( 2  \delta_\sigma^\rho \delta_\theta^\beta \delta_\phi^\delta S_u(-x)_{\alpha \omega}
+ \delta_\sigma^\rho \delta_\theta^\delta \delta_\phi^\beta S_u(-x)_{\alpha \omega} 
-\delta_\sigma^\rho \delta_\theta^\alpha \delta_\phi^\beta S_u(-x)_{\delta \omega}\Big)
    \,\langle 0|\epsilon^{abc} u_{\sigma}^a(x) d_{\theta}^b(0) s_{\phi}^c(0)|\Lambda(p)\rangle\nonumber\\
&+\Big(
2  \delta_\sigma^\beta \delta_\theta^\rho \delta_\phi^\delta S_d(-x)_{\beta \omega}
+ \delta_\sigma^\alpha \delta_\theta^\rho \delta_\phi^\beta S_d(-x)_{\delta \omega}
-\delta_\sigma^\delta \delta_\theta^\rho \delta_\phi^\beta S_d(-x)_{\alpha \omega}
\Big) \,\langle 0|\epsilon^{abc} u_{\sigma}^a(0) d_{\theta}^b(x) s_{\phi}^c(0)|\Lambda(p)\rangle
\bigg\}
   \Bigg],
\end{align}

\end{widetext}
 for $\Lambda$-$\Lambda$ transition and 

 \begin{widetext}
 
\begin{align}\label{QCD4}
	\Pi_{\mu\nu}^{QCD}(p,q)&=-\frac{\sqrt{2}}{2}\int d^4 x e^{iqx}\Bigg[\bigg\{ (\gamma_5)_{\gamma\delta}\, C_{\alpha\beta}\, ( \overleftrightarrow{D}_\mu (x)\gamma_\nu)_{\omega \rho} 
	+
	t\, (I)_{\gamma\delta}\,(C \gamma_5)_{\alpha\beta}\,( \overleftrightarrow{D}_\mu(x) \gamma_\nu)_{\omega \rho}\nonumber\\
	&+
	(\gamma_5)_{\gamma\delta} \, C_{\alpha\beta} \, ( \overleftrightarrow{D}_\nu (x)\gamma_\mu)_{\omega \rho} 
    +
	t\,(I)_{\gamma\delta}\,(C \gamma_5)_{\alpha\beta}\, ( \overleftrightarrow{D}_\nu (x)\gamma_\mu)_{\omega \rho}
	\bigg\}\nonumber\\
   & \times	\bigg\{
  \Big(   \delta_\sigma^\rho \delta_\theta^\delta \delta_\phi^\beta S_u(-x)_{\alpha \omega}
+ \delta_\sigma^\rho \delta_\theta^\alpha \delta_\phi^\beta S_u(-x)_{\delta \omega} 
\Big)
    \,\langle 0|\epsilon^{abc} u_{\sigma}^a(x) d_{\theta}^b(0) s_{\phi}^c(0)|\Lambda(p)\rangle\nonumber\\
&+\Big(
 \delta_\sigma^\alpha \delta_\theta^\rho \delta_\phi^\beta S_d(-x)_{\delta \omega}
+ \delta_\sigma^\delta \delta_\theta^\rho \delta_\phi^\beta S_d(-x)_{\alpha \omega}
\Big) \,\langle 0|\epsilon^{abc} u_{\sigma}^a(0) d_{\theta}^b(x) s_{\phi}^c(0)|\Lambda(p)\rangle
\bigg\}
   \Bigg],
\end{align}

\end{widetext}
for $\Sigma^0$-$\Lambda$ transition. Here $S_q(x)$ is the light-quark propagator  and it is given as
 \begin{widetext}
\begin{align}
\label{edmn09}
S_{q}(x) &= 
\frac{1}{2 \pi^2 x^2}\Big( i \frac{{\xslash}}{x^{2}}-\frac{m_{q}}{2 } \Big)
- \frac{\langle  \bar qq \rangle }{12} \Big(1-i\frac{m_{q} \xslash}{4}   \Big)
- \frac{\langle \bar q \sigma.G q \rangle }{192}x^2  \Big(1-i\frac{m_{q} \xslash}{6}   \Big)
-\frac {i g_s }{32 \pi^2 x^2} ~G^{\mu \nu} (x) \bigg[\rlap/{x}
\sigma_{\mu \nu} +  \sigma_{\mu \nu} \rlap/{x}
 \bigg].
\end{align}
\end{widetext}
The $\langle 0| \epsilon^{abc} u_{\sigma}^a(a_1 x) u_{\theta}^b(a_2 x) d_{\phi}^c(a_3 x)|\Sigma(p)\rangle$,
$\langle 0| \epsilon^{abc} s_{\sigma}^a(a_1 x) s_{\theta}^b(a_2 x) u_{\phi}^c(a_3 x)|\Xi(p)\rangle$  and 
$\langle 0| \epsilon^{abc} u_{\sigma}^a(a_1 x) d_{\theta}^b(a_2 x) s_{\phi}^c(a_3 x)|\Lambda(p)\rangle$ 
matrix elements are the expressions including the distribution amplitudes of the hyperon states and they are necessary for further evaluations. 
The  explicit form of this matrix elements in terms of  DAs
together with the explicit forms of DAs for hyperons are presented in the Refs.~\cite{Liu:2008yg, Liu:2009uc}.
The shape parameters of DAs of hyperons are calculated in the framework of lattice QCD \cite{Bali:2019ecy}, baryon chiral perturbation theory \cite{Wein:2015oqa} and  QCD sum rules~\cite{Liu:2008yg, Liu:2009uc}.
After using the explicit forms of the above matrix elements and the light-quark propagator, we obtain an expressions in x-space. The next step is to perform the Fourier integrals. These procedures are standard in the method used but very lengthy. Thus, we do not present these steps here.  

The desired sum rules are acquired by matching both representations of the correlation function. To do this, different and independent Lorentz structures are needed to be chosen. For this purpose, we choose $p_\mu q_\nu$, $p_\mu \gamma_\nu \qslash$, $q_\mu q_\nu$, $g_{\mu\nu}$ structures for $A^{H-H}(Q^2)$, $J^{H-H}(Q^2)$, $D^{H-H}(Q^2)$ and $\bar c^{H-H}(Q^2)$ form factors, respectively. As a result, we obtain the sum rules, 

\begin{align}\label{sigres}
\frac{\lambda_\Sigma}{m_{\Sigma}^2-p'^2} A^{\Sigma-\Sigma}(Q^2) &= - \Pi_1^{QCD}(Q^2),~~~~~~~~
\frac{\lambda_\Sigma}{m_{\Sigma}^2-p'^2} J^{\Sigma-\Sigma}(Q^2) =  \Pi_2^{QCD}(Q^2),\nonumber\\
\frac{\lambda_\Sigma}{m_{\Sigma}^2-p'^2} D^{\Sigma-\Sigma}(Q^2) &= 2\, \Pi_3^{QCD}(Q^2),~~~~~~~~
\frac{\lambda_\Sigma}{m_{\Sigma}^2-p'^2} \bar c^{\Sigma-\Sigma}(Q^2) =  \frac{1}{2 m^2_{\Sigma}}\Pi_4^{QCD}(Q^2),
\end{align}

\begin{align}
\frac{\lambda_\Xi}{m_{\Xi}^2-p'^2} A^{\Xi-\Xi}(Q^2) &= - \Pi_5^{QCD}(Q^2),~~~~~~~~
\frac{\lambda_\Xi}{m_{\Xi}^2-p'^2} J^{\Xi-\Xi}(Q^2) =  \Pi_6^{QCD}(Q^2),\nonumber\\
\frac{\lambda_\Xi}{m_{\Xi}^2-p'^2} D^{\Xi-\Xi}(Q^2) &= 2\, \Pi_7^{QCD}(Q^2),~~~~~~~~
\frac{\lambda_\Xi}{m_{\Xi}^2-p'^2} \bar c^{\Xi-\Xi}(Q^2) =  \frac{1}{2 m^2_{\Xi}}\Pi_8^{QCD}(Q^2),
\end{align}

\begin{align}
\frac{\lambda_\Lambda}{m_{\Lambda}^2-p'^2} A^{\Lambda-\Lambda}(Q^2) &= - \Pi_9^{QCD}(Q^2),~~~~~~~~
\frac{\lambda_\Lambda}{m_{\Lambda}^2-p'^2} J^{\Lambda-\Lambda}(Q^2) =  \Pi_{10}^{QCD}(Q^2),\nonumber\\
\frac{\lambda_\Lambda}{m_{\Lambda}^2-p'^2} D^{\Lambda-\Lambda}(Q^2) &= 2\, \Pi_{11}^{QCD}(Q^2),~~~~~~~~
\frac{\lambda_\Lambda}{m_{\Lambda}^2-p'^2} \bar c^{\Lambda-\Lambda}(Q^2) =  
\frac{1}{2 m^2_{\Lambda}}\Pi_{12}^{QCD}(Q^2),
\end{align}

\begin{align}\label{siglamres}
\frac{\lambda_{\Sigma^0}}{m_{\Sigma^0}^2-p'^2} A^{\Sigma^0-\Lambda}(Q^2) &=
- \Pi_{13}^{QCD}(Q^2),~~~~~~~~
\frac{\lambda_{\Sigma^0}}{m_{\Sigma^0}^2-p'^2} J^{\Sigma^0-\Lambda}(Q^2) =
 \Pi_{14}^{QCD}(Q^2),\nonumber\\
\frac{\lambda_{\Sigma^0}}{m_{\Sigma^0}^2-p'^2} D^{\Sigma^0-\Lambda}(Q^2) &= 2\,
\Pi_{15}^{QCD}(Q^2),
~~~~~~~~
\frac{\lambda_{\Sigma^0}}{m_{\Sigma^0}^2-p'^2} \bar c^{\Sigma^0-\Lambda}(Q^2) =  
\frac{2}{( m_{\Lambda}+ m_{\Sigma^0})^2}\Pi_{16}^{QCD}(Q^2).
\end{align}

The  $\Pi_{i}^{QCD}(Q^2)$  functions  appear in Eqs. (\ref{sigres}) to (\ref{siglamres}) are very lengthy. Hence, as an example,  we present  the result of the   $\Pi_{1}^{QCD}(Q^2)$:
\begin{eqnarray}
 \Pi_{1}^{QCD}(Q^2) &=&\frac{m_{\Sigma}}{4}\int_0^1 dx_2 \int_0^{1-x_2} dx_1 \frac{x_2}{(q- px_2)^2}\bigg[\big (1-t\big)\big(3A_1+6A_3-V_1+2V_3)+2\,(1+t)(P_1-S_1-T_1
\nonumber\\
\nonumber\\
&&+2T_7\big)\bigg] (x_1,x_2,1-x_1-x_2)
\nonumber\\
\nonumber\\
 & &+\frac{m_{\Sigma}}{4}\int_0^1 d\alpha \int_\alpha^1 dx_2 \int_0^{1-x_2} dx_1  \frac{1}{(q- p \alpha)^2}
\bigg [2\,\big(1-t\big)\big(2A_1-2A_2+2A_3-V_1+V_2+V_3\big)
 \nonumber\\
\nonumber\\
& &+\big (1+t\big)\big(7T_1+4 T_2+3 T_3+11 T_7\big)\bigg](x_1,x_2,1-x_1-x_2)
\nonumber
\end{eqnarray}
\begin{eqnarray}
&  &+\frac{m^3_{\Sigma}}{2}\int_0^1 d\alpha \int_\alpha^1 dx_2 \int_0^{1-x_2} dx_1  \frac{\alpha^2}{(q- p \alpha)^4}\bigg[\big(1-t\big)\big(3A_1-3A_2-A_4-A_5-V_1+V_2+2V_3-V_4\big) 
\nonumber\\
\nonumber\\
& &+\big (1+t\big)\big(-P_1+P_2-S_1+S_2-2T_1- T_3+T_5+6 T_7\big)\bigg](x_1,x_2,1-x_1-x_2)
\nonumber\\
\nonumber\\
&  &+\frac{m^3_{\Sigma}}{4}\int_0^1 d\beta \int_\beta^1 d_\alpha \int_\alpha^1 dx_2 \int_0^{1-x_2} dx_1  \frac{\beta}{(q- p \beta)^4} \bigg[4\,\big(1-t\big)\big(2A_1-2A_2+2A_3+2A_4-2A_5+2A_6
\nonumber\\
\nonumber\\
 &&-V_1 +V_2+V_3+V_4+V_5-V_6\big)+ \big(1+t\big)\big(14T_1+9T_2+5T_3+5T_4+9T_5-14T_6+23T_7
\nonumber\\
\nonumber\\
&&+23T_8\big)\bigg](x_1,x_2,1-x_1-x_2)
\nonumber\\
\nonumber\\
  &&+2 m^5_{\Sigma}\int_0^1 d\beta \int_\beta^1 d_\alpha \int_\alpha^1 dx_2 \int_0^{1-x_2} dx_1  \frac{\beta^3}{(q- p \beta)^6} \bigg[ \big(1+t\big)\big(T_2-T_3-T_4+T_5+T_7+T_8\big)\bigg]
\nonumber\\
\nonumber\\
&&\times(x_1,x_2,1-x_1-x_2).
\end{eqnarray}
Here, $ S_i $, $ P_i $, $ A_i $,  $ V_i $ and  $ T_i $ are DAs of different twists. They are expressed in terms of hyperon's wavefunctions. The wavefunctions are also functions of different parameters that all are presented in Refs.~\cite{Liu:2008yg, Liu:2009uc}.
In order to remove the unwanted contributions corresponding to the excited and continuum states,
the Borel transformation and continuum subtraction are applied. 
Following these  procedures, the contributions of the higher states and continuum are exponentially suppressed. 
The Borel transformation and the continuum subtraction are carried out
by employing the subsequent replacement  rules \cite{Braun:2006hz}:
\begin{align}
		\int dz \frac{\rho(z)}{\Delta^2}\rightarrow &-\int_{x_0}^1\frac{dz}{z}\rho(z) e^{-s(z)/M^2},
		 \nonumber\\
		\int dz \frac{\rho(z)}{\Delta^4}\rightarrow & \frac{1}{M^2} \int_{x_0}^1\frac{dz}{z^2}\rho(z) e^{-s(z)/M^2}
		+\frac{\rho(x_0)}{Q^2+x_0^2 m^2_{H}} e^{-s_0/M^2},
\nonumber\\
		\int dz \frac{\rho(z)}{\Delta^6}\rightarrow &-\frac{1}{2M^4}\int_{x_0}^1\frac{dz}{z^3}\rho(z) e^{-s(z)/M^2}
		-\frac{1}{2M^2}\frac{\rho(x_0)}{x_0(Q^2+x_0^2m_H^2)}e^{-s_0/M^2}
\nonumber\\
		&+\frac{1}{2}\frac{x_0^2}{Q^2+x_0^2m_H^2}\bigg[\frac{d}{dx_0}\frac{\rho(x_0)}{x_0(Q^2+x_0^2m_H^2)}\bigg]e^{-s_0/M^2},
	\label{subtract3}
\end{align}
where,
\begin{align}
\Delta =& q-zp,\nonumber\\
s(z) =& (1-z)m^2_{H}+\frac{1-z}{z}Q^2,\nonumber\\
x_0 =& \Big(\sqrt{(Q^2+s_0-m^2_{H})^2+4 m^2_{H} Q^2}-(Q^2+s_0-m^2_{H})\Big)/{2m^2_{H}}.
\end{align}

\section{Numerical Results}\label{secIII}

This section is dedicated to the numerical analysis for gravitational form factors of hyperon states.
 The distribution amplitudes of 
 the $\Sigma$, $\Xi$ and $\Lambda$ baryons 
 have been evaluated  in the framework of   QCD sum rule in Refs.~\cite{Liu:2009uc, Liu:2008yg}.
In Table \ref{parameter_table}, we present the numerical values of the input parameters inside the
DAs of hyperons from these studies, which are used in the numerical computations. 
 For the numerical computations, the hyperon masses are also taken as,
$m_{\Sigma} = 1189.37 \pm 0.07$ MeV, $m_{\Xi} =1314.86 \pm 0.20$ MeV, $m_{\Sigma^0} = 1192.642 \pm 0.024$ MeV and $m_{\Lambda} = 1115.685 \pm 0.006 $ MeV 
\cite{Tanabashi:2018oca}. 
 Additionally, we need the values of the residues of $\Sigma$, $\Xi$, $\Sigma^0$ and $\Lambda$ baryons, which are borrowed from \cite{Aliev:2002ra}.
\begin{table}[t]
	\addtolength{\tabcolsep}{2pt}
	\begin{center}
\begin{tabular}{ |l|l|l|l| }
\hline
\multicolumn{3}{ |c| }{ Parameters in wavefunctions of hyperons.} \\
\hline\hline
$~~~~~~~~~~~~~~~~~~~~\Sigma$ & $~~~~~~~~~~~~~~~~~~~~\Xi $& $~~~~~~~~~~~~~~~~~~~~\Lambda$ \\ \hline\hline
\multirow{4}{*}{} &  &  \\
 $f = (9.4 \pm 0.4)\times 10^{-3}$~GeV$^2$  & $f  = (9.9 \pm 0.4) \times10^{-3}~GeV^2$      & $f  = (6.0 \pm 0.3) \times 10^{-3}~GeV^2$        \\
 $\lambda_1= (-2.5 \pm 0.1)\times  10^{-2}$~GeV$^2$ & $\lambda_1 = (-2.8 \pm 0.1) \times10^{-2}~GeV^2$  &$\lambda_1 = (1.0 \pm 0.3) \times 10^{-2}~GeV^2$   \\
 $\lambda_2 = (4.4 \pm 0.1) \times  10^{-2}$~GeV$^2$ & $\lambda_2 = (5.2 \pm 0.2)10^{-2} \times ~GeV^2$  & $\lambda_2 = (0.83\pm 0.05) \times10^{-2}~GeV^2$  \\ 
$\lambda_3 = (2.0  \pm 0.1) \times  10^{-2}$~GeV$^2$  & $\lambda_3 = (0.17 \pm 0.1) \times 10^{-2}~GeV^2$
 & $\lambda_3 = (0.83\pm 0.05) \times 10^{-2}~GeV^2$  
\\ \hline\hline
\end{tabular}
\caption{The numerical values of the parameters   in the wavefunctions of hyperons taken from Refs. \cite{Liu:2009uc, Liu:2008yg}.
          }
          \label{parameter_table}
         \end{center}
\end{table}

The estimations for the gravitational form factors of hyperon states depend on several auxiliary parameters: mixing parameter $t$, the Borel mass parameter $M^2$ and  continuum threshold $s_0$.
Based on the standard prescriptions of the approach used, the  physical observables should weakly depend on these auxiliary parameters.
The arbitrary mixing parameter $t$ is chosen such that, the predictions of the gravitational form factors are practically independent of the values of $t$. 
From the numerical analyzes, it is seen that in the region -0.19 $\leq cos\theta \leq$ -0.44  the gravitational form factors weakly depend on $t$, where   tan$\theta$ = $t$.
The working interval for the continuum threshold $s_0$ is obtained considering the fact that the gravitational form factors  also weakly depend on this parameter. It is not totally arbitrary, but depend on the energy of the first excited states in the channels under consideration. 
We follow the steps described below to get the working window of Borel mass parameter $ M^2 $.
The lower limit of $M^2$  is obtained requiring that the perturbative contribution exceeds over the nonperturbative one and the series of nonperturbative operators are convergent.
The upper limit of $M^2$ is obtained  by the requirement that the contributions of continuum and  higher states should be less than the ground state contribution.

Performed numerical calculations show that the regions
\begin{eqnarray}
& &2.30~\mbox{GeV}^2 \leq s_0 \leq 3.00~\mbox{GeV}^2~\mbox{for}~\Sigma-\Sigma, \nonumber\\
& &2.90~\mbox{GeV}^2 \leq s_0 \leq 3.60~\mbox{GeV}^2 ~\mbox{for}~ \Xi-\Xi,\nonumber\\
& &2.00~\mbox{GeV}^2 \leq s_0 \leq 2.60~\mbox{GeV}^2~\mbox{for}~ \Lambda-\Lambda,\nonumber\\
& &2.30~\mbox{GeV}^2 \leq s_0 \leq 3.00~\mbox{GeV}^2  ~\mbox{for}~ \Sigma^0-\Lambda,
\end{eqnarray}
\begin{eqnarray}
& &2.00~\mbox{GeV}^2 \leq M^2 \leq 3.00~\mbox{GeV}^2~\mbox{for}~\Sigma-\Sigma, \nonumber\\
& &2.50~\mbox{GeV}^2 \leq M^2 \leq 3.50~\mbox{GeV}^2 ~\mbox{for}~ \Xi-\Xi, \nonumber\\
& &2.00~\mbox{GeV}^2 \leq M^2 \leq 3.00~\mbox{GeV}^2~\mbox{for}~ \Lambda-\Lambda, \nonumber\\
& &2.00~\mbox{GeV}^2 \leq M^2 \leq 3.00~\mbox{GeV}^2  ~\mbox{for}~ \Sigma^0-\Lambda,
\end{eqnarray}
satisfy all the aforementioned constraints on $ M^2$ and $s_0$.
To visualize effects of $t$, $M^2$ and $s_0$ on the gravitational form factors we depict the dependence of these form factors on the auxiliary parameters in Fig. \ref{hyperonMsqfigs}. As it can be seen from this figure, 
the gravitational form factors are quite stable against the  variation of these parameters in their working windows.
Thus the selected working intervals for the auxiliary parameters satisfy the criteria of the method.
In Fig. \ref{hyperonQsqfigs} we present dependence of the gravitational form factors $ A^{H-H}(Q^2)$, $ J^{H-H}(Q^2)$, $ D^{H-H}(Q^2)$ and $ \bar c^{H-H}(Q^2)$ on $Q^2$ in the interval 1.0 GeV$^2 \leq Q^2 \leq $ 10.0 GeV$^2$ at fixed values of auxiliary parameters, using the central values of all input parameters in DAs. From this figure,  we see that the $Q^2$ dependencies of the gravitational form factors are smoothly varying and decrease with increasing of the $ Q^2 $ as expected.
The only exception is the form factor $ \bar c^{\Sigma-\Sigma}(Q^2)$. The behaviour of this form factor is not reliable. Therefore, we will not present the fit result of this form factor.
 Our numerical results for the transitional gravitational form factors of $ \Sigma ^ 0- \Lambda $ transition are very small and close to zero, hence, we do not show  the $Q^2$ and  $M^2$  dependence of the form factors corresponding to this transition. 
%

To extrapolate the gravitational form factors to low-momentum region, $Q^2 =0$, we fit the light-cone QCD sum rule results to different forms of mono-pole and dipole, which unluckily fail to give  reasonable characterizations of data with a two-parameter fit. 
However,  our numerical studies indicate that the gravitational form factors of  hyperons are nicely characterized employing the multipole fit function,
 \begin{align}
 F^{H-H}(Q^2)= \frac{ F^{H-H}(0)}{\Big(1+ Q^2\,m_{p}\Big)^p},
\end{align} 
where $F^{H-H}(Q^2)$= $ A^{H-H}(Q^2)$, $ J^{H-H}(Q^2)$, $ D^{H-H}(Q^2)$ and $ \bar c^{H-H}(Q^2)$ and;
$F^{H-H}(0)$ = $ A^{H-H}(Q^2=0)$, $ J^{H-H}(Q^2=0)$, $ D^{H-H}(Q^2=0)$ and $ \bar c^{H-H}(Q^2=0)$.
We should say here that the light-cone QCD sum rule is reliable only at $Q^2 \geq 1.0$ GeV$^2$.
 Our numerical values for the form factors at $Q^2 =0$, obtained using the shape parameters from the QCD sum rules,  are presented in Table \ref{fit_table}.

To our best knowledge, this is the first study in the literature dedicated to the investigation of the hyperon's gravitational  form factors.
Therefore,  experimental data or theoretical estimations are not yet available to compare them with our numerical results.
However, we may compare these results with the nucleon's gravitational form factors especially for the form factor  $D^{H-H}(Q^2)$. 
Various estimations for the $D^{N-N}(0)$ are as follows. Lattice QCD predictions:  $D^{N-N}(0) = -1.76 \pm 0.09$ \cite{Bratt:2010jn}, $D^{N-N}(0) = -2.27 \pm 0.30$ \cite{Hagler:2007xi}, 
chiral perturbation theory  estimates:  $D^{N-N}(0) = -1.93 \pm 0.06$ \cite{Dorati:2007bk},  
light-cone QCD sum rule results: $D^{N-N}(0) = -2.63 \pm 0.22$ \cite{Anikin:2019kwi},
 $D^{N-N}(0) = -2.29 \pm 0.58$ (set-I) \cite{Azizi:2019ytx},
 $D^{N-N}(0) = -2.05 \pm 0.40$ (set-II) \cite{Azizi:2019ytx}, KM15 global fit prediction: $D^{N-N}(0) = -2.18 \pm 0.21$  \cite{Anikin:2017fwu} and JLab data:  $D^{N-N}(0) = -2.11 \pm 0.46$ \cite{Burkert:2018bqq}. 
 It should  be noted here that the results of other theoretical approaches have been obtained at different renormalization scale and these results have been shifted to the renormalization scale $\mu^2 = 1$ GeV$^2$ (For details, see e.g.,  \cite{Azizi:2019ytx}).
%
As one can see from these predictions, the numerical results for the D-term of hyperons  obtained in the present work are close to the nucleon's D-term and we see a reasonable SU(3) flavor violation.  

\begin{table}[t]
	\addtolength{\tabcolsep}{2pt}
	\begin{center}
\begin{tabular}{c|c|c|cccc}
				\hline\hline
		Transition~~~~~~ &~~~~~~  $F^{H-H}(0)$ ~~~~~~ & ~~~~~~ $m_p (GeV^{-2})$ ~~~~~~&~~~~~~ $p$ ~~~~~~ \\[0.5ex]
		 \hline\hline

		                            &$A(0) = 0.86 \pm 0.12$ & $1.15 \pm 0.08$ & $3.2 \pm 0.2$   \\
		$\Sigma-\Sigma$     &$J(0) = 0.40 \pm 0.06$ & $1.16 \pm 0.10$ & $3.2 \pm 0.2$  \\
		                            &$D(0) = -2.65 \pm 0.25$ & $1.13 \pm 0.07$ & $3.2 \pm 0.1$  \\
                                     &$\bar c(0) = -$ & $-$ & $-$ \\
		 \hline\hline
				                   &$A(0) = 0.53 \pm 0.05$ & $1.10 \pm 0.08$ & $3.2 \pm 0.2$   \\
		$\Xi-\Xi$                 &$J(0) = 0.25 \pm 0.02$ & $0.93 \pm 0.06$ & $3.2 \pm 0.2$  \\
		                            &$D(0) = -2.30 \pm 0.18$ & $1.02 \pm 0.06$ & $3.2 \pm 0.1$  \\
                                     &$\bar c(0) =-0.12 \pm 0.01$ & $0.98 \pm 0.06$ & $3.8 \pm 0.2$ \\
		 \hline\hline
			                            &$A(0) = 0.58 \pm 0.10$ & $1.10 \pm 0.10$ & $3.2 \pm 0.2$   \\
		$\Lambda-\Lambda$     &$J(0) = 0.32 \pm 0.06$ & $1.20 \pm 0.10$ & $3.2 \pm 0.2$  \\
		                                &$D(0) = -2.53 \pm 0.12$ & $1.10 \pm 0.05$ & $3.2 \pm 0.1$  \\
                                         &$\bar c (0) = -0.10 \pm 0.01$ & $1.10 \pm 0.10$ & $3.8 \pm 0.2$ \\
		\hline\hline
	\end{tabular}
\caption{The values of form factors at $Q^2 =0$ obtained using the multipole fit parameterization and the values of shape parameters from QCD sum rules.}
	\label{fit_table}
		\end{center}
\end{table}

\begin{table}[t]
	\addtolength{\tabcolsep}{2pt}
	\begin{center}
\begin{tabular}{c|c|c|cccc}
				\hline\hline
		Transition~~~~~~ &~~~~~~  $F^{H-H}(0)$ ~~~~~~ & ~~~~~~ $m_p (GeV^{-2})$ ~~~~~~&~~~~~~ $p$ ~~~~~~ \\[0.5ex]
		 \hline\hline

		                            &$A(0) = 0.98 \pm 0.18$ & $1.13 \pm 0.08$ &  $3.2 \pm 0.2$   \\
		$\Sigma-\Sigma$             &$J(0) = 0.44 \pm 0.05$ & $1.09 \pm 0.09$ &  $3.2 \pm 0.2$  \\
		                            &$D(0) = -2.90 \pm 0.40$ &$1.09 \pm 0.07$ & $3.2 \pm 0.1$  \\
                                     &$\bar c(0) = -$ & $-$ & $-$ \\
		 \hline\hline
				                    &$A(0) = 0.70 \pm 0.07$ & $1.16 \pm 0.08$ & $3.2 \pm 0.2$   \\
		$\Xi-\Xi$                   &$J(0) = 0.22 \pm 0.04$ & $0.93 \pm 0.05$ & $3.2 \pm 0.2$  \\
		                            &$D(0) = -2.85 \pm 0.30$ & $1.0 \pm 0.05$ & $3.2 \pm 0.1$  \\
                                    &$\bar c(0) =-0.13 \pm 0.02$ & $0.95 \pm 0.07$ & $3.8 \pm 0.2$ \\
		 \hline\hline
			                        &$A(0) = 0.87 \pm 0.17$ & $1.22 \pm 0.10$ & $3.2 \pm 0.2$   \\
		$\Lambda-\Lambda$           &$J(0) = 0.42 \pm 0.06$ & $1.20 \pm 0.17$ & $3.2 \pm 0.2$  \\
		                            &$D(0) = -2.70 \pm 0.50$& $1.15 \pm 0.10$ & $3.2 \pm 0.1$  \\
                                    &$\bar c (0) = -0.11 \pm 0.01$ & $1.13 \pm 0.11$ & $3.8 \pm 0.2$ \\
		\hline\hline
	\end{tabular}
\caption{The values of form factors at $Q^2 =0$ obtained using the multipole fit parameterization and the values of shape parameters from  lattice QCD.}
	\label{fit_table-1}
		\end{center}
\end{table}

As we previously mentioned, the shape parameters of DAs of hyperons are now available from  lattice QCD \cite{Bali:2019ecy}, as well. Using these values by changing the scale of the parameters to the one used in the present study, we obtain the values of the form factors at $Q^2 =0$ as presented in Table \ref{fit_table-1}. Comparing the values of the form factors in this table with those in Table \ref{fit_table}, we see that the central values differ from each other by 
$ 8\%-34\% $. These amounts of the changes are reasonable as the DAs of the baryons generally have strong dependencies to the shape parameters.

After obtaining the gravitational form factors, we can use them to calculate  some mechanical properties associated with the hyperons such as mechanical radius square ($\langle r^2_{\text{mech}}\rangle$) as well as energy ($\cal E$) and pressure ($p_0$) distributions at the center of these particles. The related formulas  are given as \cite{Polyakov:2018zvc},
\begin{eqnarray}
\label{Pres}
&&p_0  =-\frac{1}{24\,\pi^2\, m_H} \int^{\infty}_{0} dy \,y\,\sqrt{y}\, [ D(y)-\bar c (y) ],\\
&&{\cal E}=\frac{m_H}{4\,\pi^2} \int^{\infty}_{0} dy \,\sqrt{y}\,\Big[ A(y) + \frac{y}{4m^2_H} [A(y) -2J(y)+ D(y)+ \bar c (y)]\Big],\\
&& \langle r^2_{\text{mech}}\rangle = \frac {6\, D(0)}{\int^{\infty}_{0} dy\, D(y)},
\end{eqnarray}
where $y = Q^2$.

\begin{table}[htp]
	\addtolength{\tabcolsep}{2pt}
	\begin{center}
\begin{tabular}{c|c|c|cccc}
				\hline\hline
		Transition &  ~~~ $p_0$  (GeV/fm$^3$)~~~&~~~ ${\cal E}$ (GeV/fm$^3$) ~~~&~~~$\langle {r^2_{\text{mech}}\rangle}$ (fm$^{2}$)~~~& \\[0.5ex]
		\hline\hline

		 $\Sigma-\Sigma$ &$  0.54 \pm 0.14$ &$ 2.41 \pm 0.63$ &$0.61 \pm 0.06$&  \\
		  \hline\hline
	           $\Xi-\Xi$  &$  0.62 \pm 0.10$ &$ 2.09 \pm 0.62 $ &$ 0.52 \pm 0.04$&  \\
		           		  \hline\hline
		    $\Lambda-\Lambda$ &$  0.59 \pm 0.15$ &$ 2.08 \pm 0.56 $ &$0.59 \pm 0.05$&  \\
		                            
		\hline\hline
	\end{tabular}
\caption{ The values of mechanical quantities for hyperons  using the shape parameters from QCD sum rules.}
	\label{mech_table}
		\end{center}
\end{table}

\begin{table}[htp]
	\addtolength{\tabcolsep}{2pt}
	\begin{center}
\begin{tabular}{c|c|c|cccc}
				\hline\hline
		Transition &  ~~~ $p_0$  (GeV/fm$^3$)~~~&~~~ ${\cal E}$ (GeV/fm$^3$) ~~~&~~~$\langle {r^2_{\text{mech}}\rangle}$ (fm$^{2}$)~~~& \\[0.5ex]
		\hline\hline

		 $\Sigma-\Sigma$ &$  0.64 \pm 0.14$ &$ 2.86 \pm 0.68$ &$0.58 \pm 0.06$&  \\
		  \hline\hline
	           $\Xi-\Xi$  &$  0.69 \pm 0.10$ &$ 2.45 \pm 0.60 $ &$ 0.54 \pm 0.04$&  \\
		           		  \hline\hline
		    $\Lambda-\Lambda$ &$  0.56 \pm 0.13$ &$ 2.23 \pm 0.51 $ &$0.62 \pm 0.08$&  \\
		                            
		\hline\hline
	\end{tabular}
\caption{ The values of mechanical quantities for hyperons using the shape parameters from lattice QCD.}
	\label{mech_table-1}
		\end{center}
\end{table}

The numerical results  of  mechanical properties of hyperons are presented in Tables \ref{mech_table} and \ref{mech_table-1}, for the shape parameters from the QCD sum rules and lattice QCD, respectively.  Comparing the central  values of the parameters from these two tables, we see that the differences among the corresponding values lie in the interval $ 4\%-16\% $.  The minimum difference belongs to the $ \langle {r^2_{\text{mech}}\rangle}_\Xi $ and the maximum difference between the values is related to the ${\cal E}_\Sigma$. Our results may be checked via different theoretical approaches.

\section{Summary and Concluding Remarks}\label{secIV}

 We calculated the quark part of the  gravitational form factors of hyperons by using the light-cone QCD sum rule method. We used the general forms of the interpolating currents for the hyperons and the quark part of the energy-momentum current together with the DAs of the hyperonic states calculated via both the QCD sum rules and lattice QCD. These form factors carry information about the angular momentum and mass as well as energy and pressure distributions inside the baryons. We obtained  that the $Q^2$ dependence of the hyperon's  gravitational form factors are nicely characterized by a multipole fit function. Using these fit functions, we obtained the static form factors, i. e., their values at $ Q^2=0 $. The hyperons under study are unstable and have a small life-time, hence measurements on their properties are very difficult compared with the nucleons. We have a good experimental information on the electromagnetic form factors of nucleons, even for the D-term of the nucleons we have some data provided by JLab. We also have a good knowledge of theory on the electromagnetic, axial, tensor and gravitational form factors of the nucleons. However, the number of works dedicated to the electromagnetic, axial and tensor form factors of hyperons are very limited. In the case of gravitational form factors there are no theory predictions are available. Hence, our results may be checked via different methods in future. We hope we can have experimental data on the gravitational form factors of hyperons in future. Comparison of different results on these form factors can help us gain valuable information on the internal structures of hyperons. Our analyses show that using the shape parameters from the QCD sum rules or lattice QCD changes the results of form factors at $ Q^2=0 $  maximally by $ 34\% $.

As a by product, we calculated some mechanical properties of hyperons such as the mean mechanical radius square as well as the  pressure and energy distributions at the center of these particles, as well. The results indicate that the $\langle {r^2_{\text{mech}}\rangle}_H$ for $ H=\Sigma, \Lambda $ are considerably bigger than that of the nucleons, $\langle {r^2_{\text{mech}}\rangle}_N=(0.52-0.54) \pm 0.05~ fm^{2}$,  obtained 
via the same method in Ref. \cite{Azizi:2019ytx}.  However, the mean redius square for $ \Xi $ hyperon is comparable with that of the nucleon. We observed that switching the values of the shape parameters from QCD sum rules to those from lattice QCD, changes the results of the mechanical parameters by  $ 4\%-16\% $.
 Our results on the mechanical properties of hyperons may be checked via different theoretical approaches as well as Lattice QCD. Any future experimental data and their comparison with our estimations can give valuable knowledge on the internal structure and geometrical shapes of hyperons. 

\section{Acknowledgements}
The work of U. \"{O}. is supported under 2218-National Postdoctoral Research Scholarship Programme by the scientific and technological research council of Turkey (TUBITAK).

\bibliography{refs}

\begin{widetext}
 
 \begin{figure}[t]
\centering
  \subfloat[]{ \includegraphics[width=0.3\textwidth]{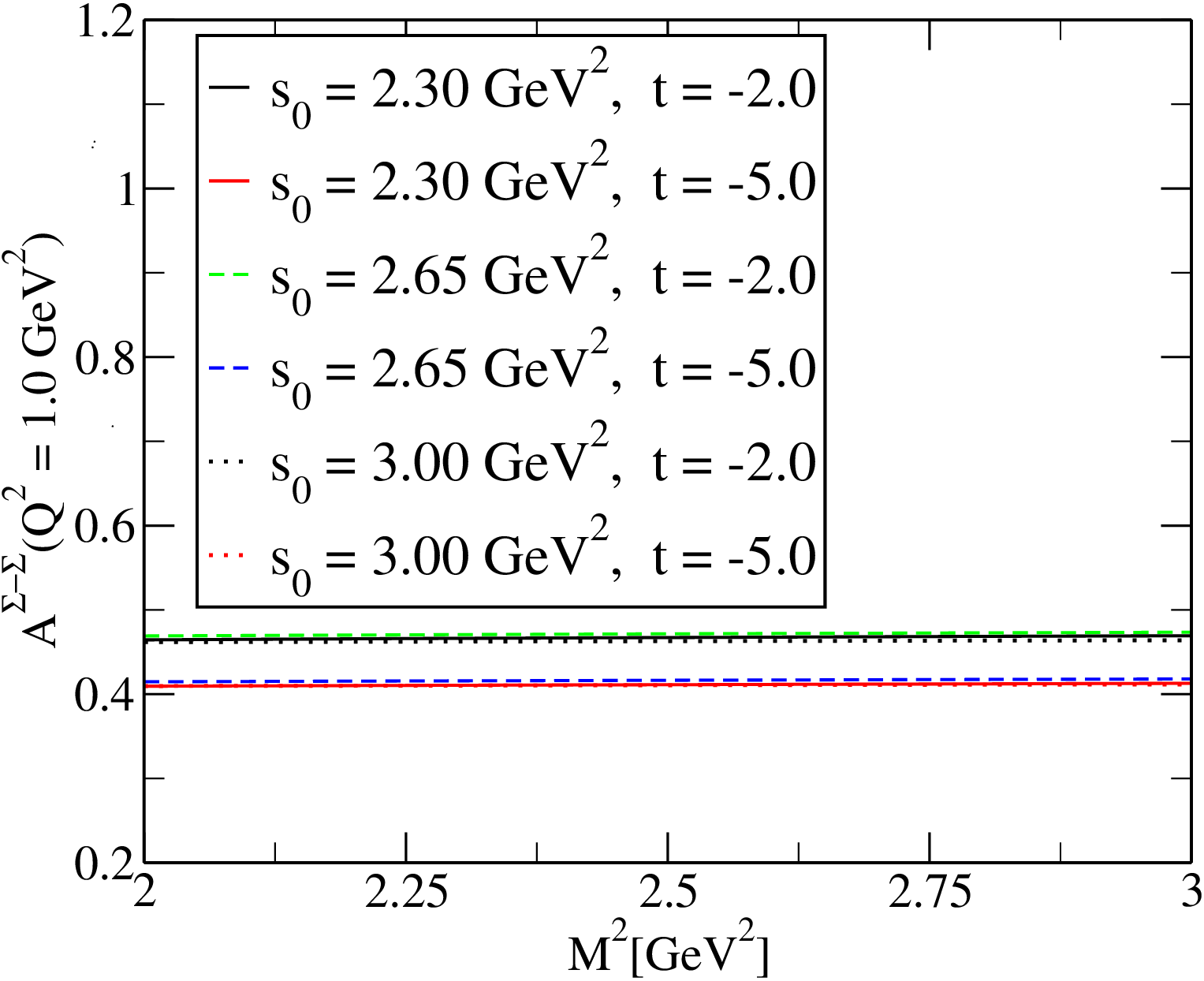}}~~~
 \subfloat[]{ \includegraphics[width=0.3\textwidth]{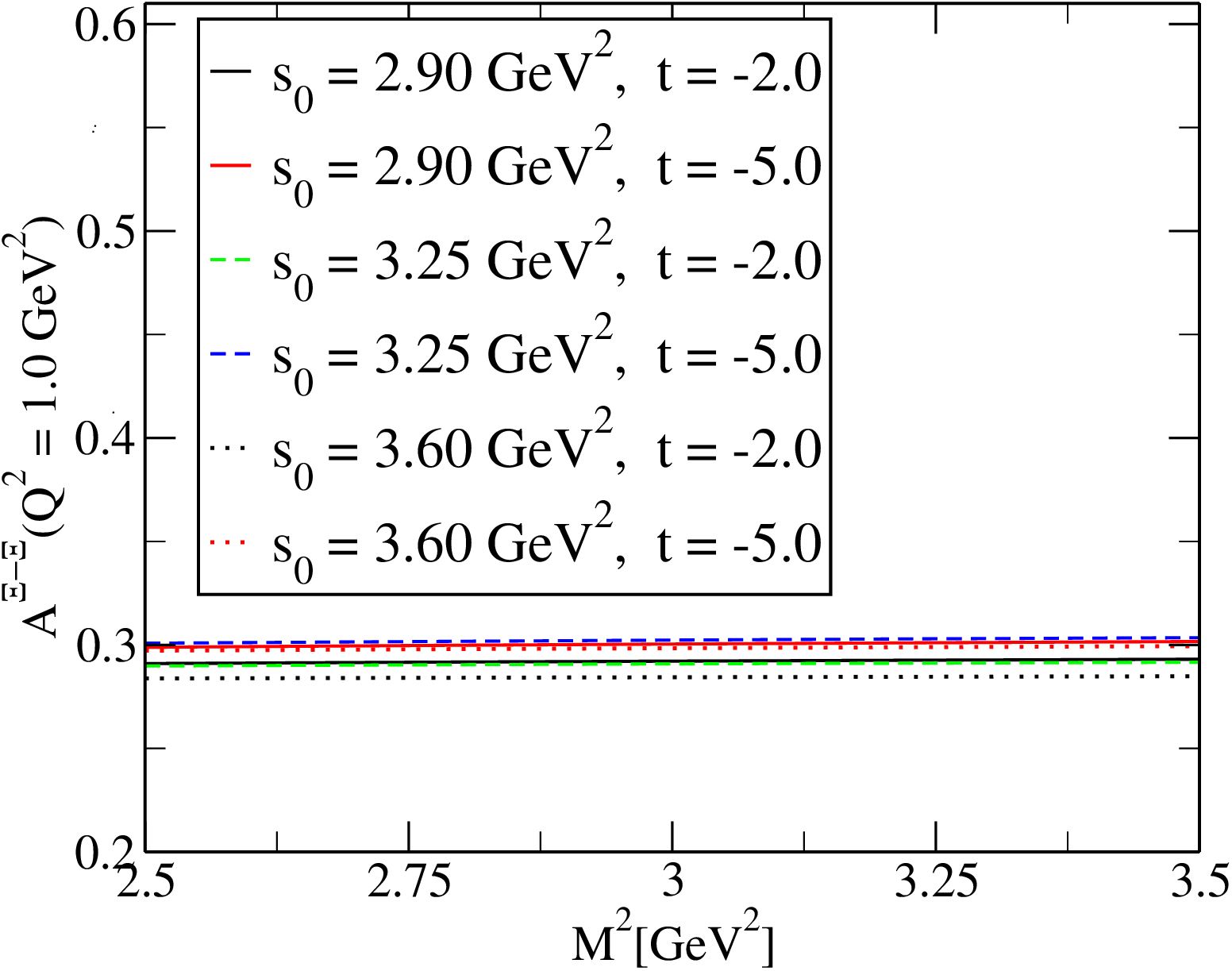}}~~~
 \subfloat[]{ \includegraphics[width=0.3\textwidth]{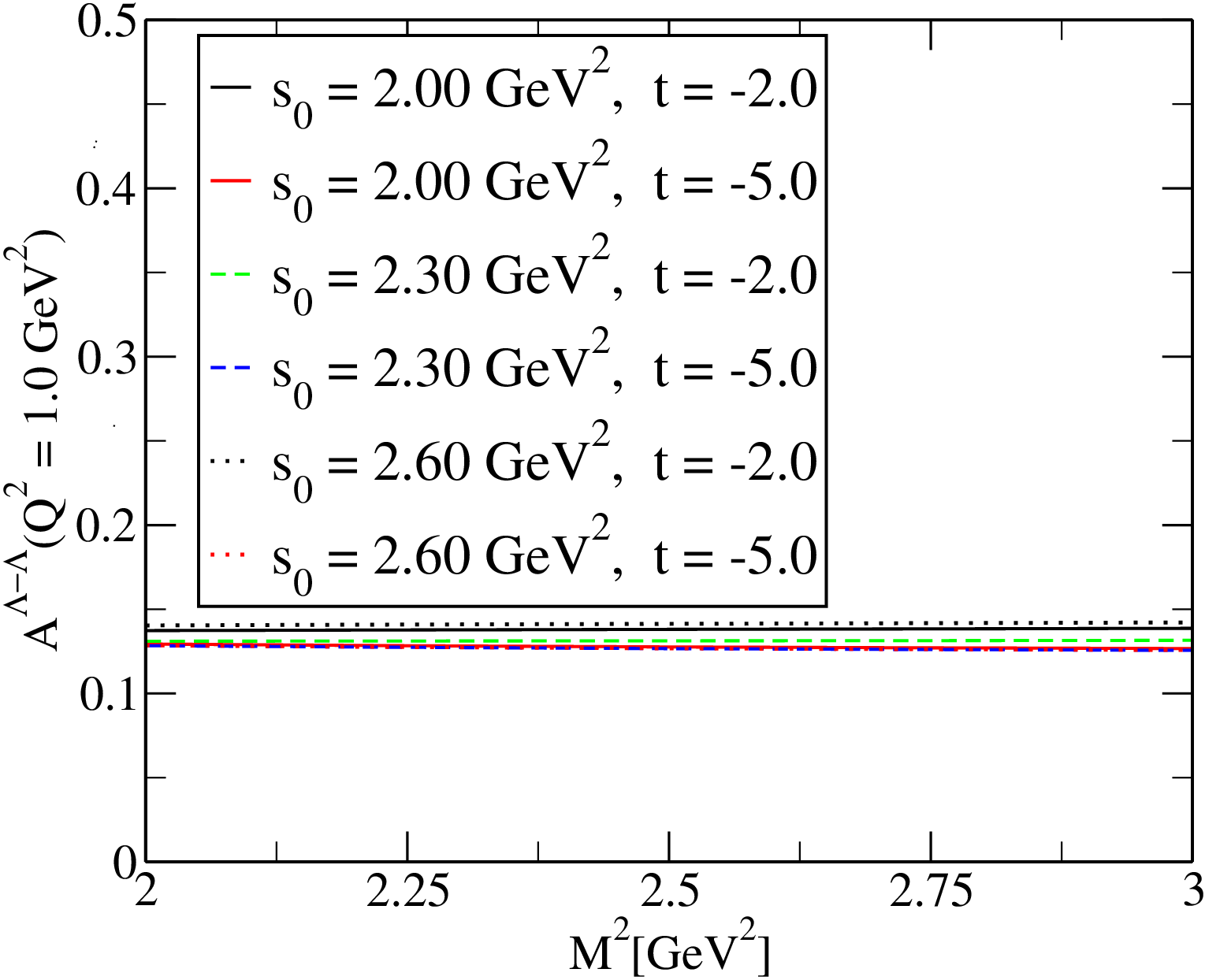}}~~~\\
 \subfloat[]{ \includegraphics[width=0.3\textwidth]{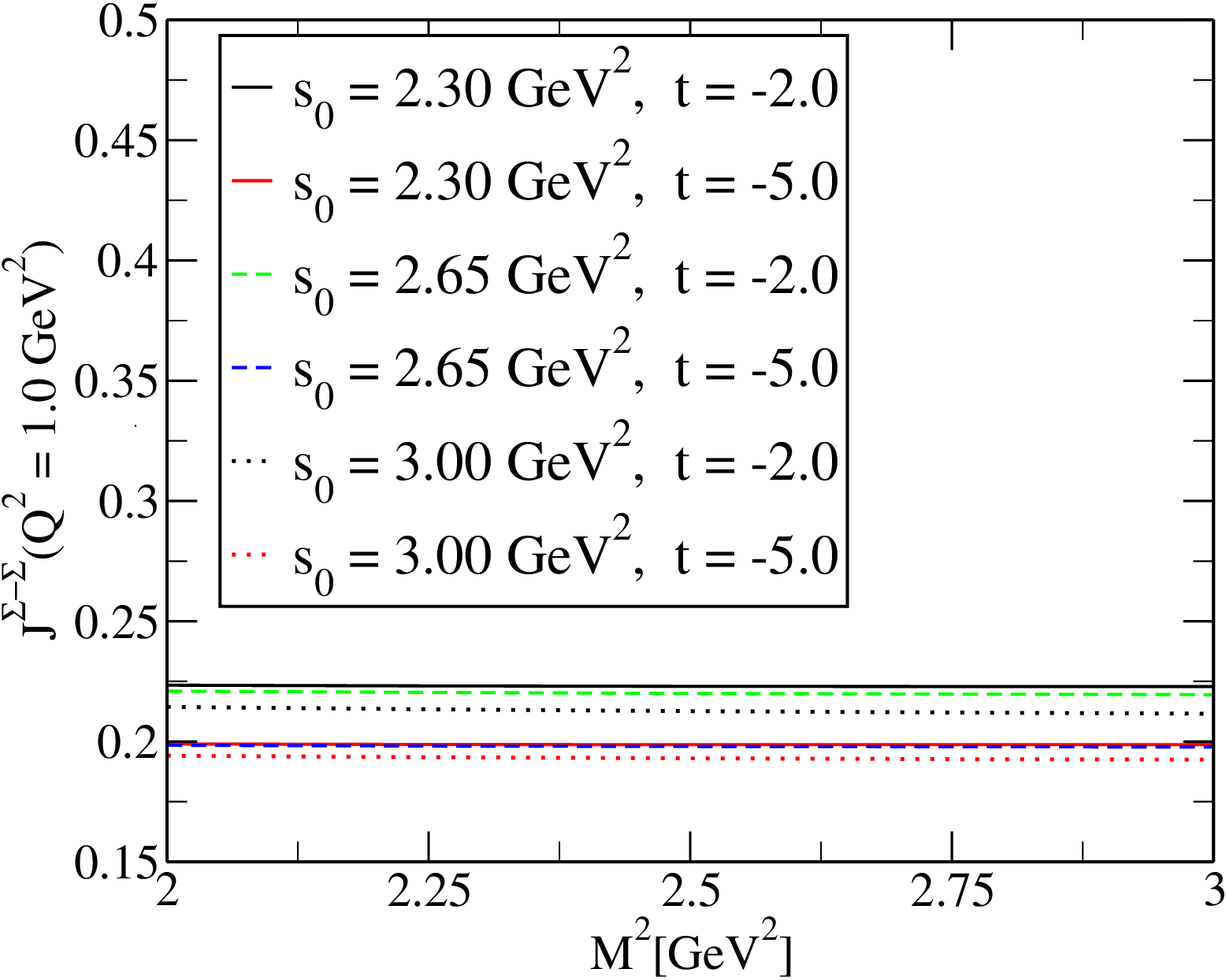}}~~~
\subfloat[]{ \includegraphics[width=0.3\textwidth]{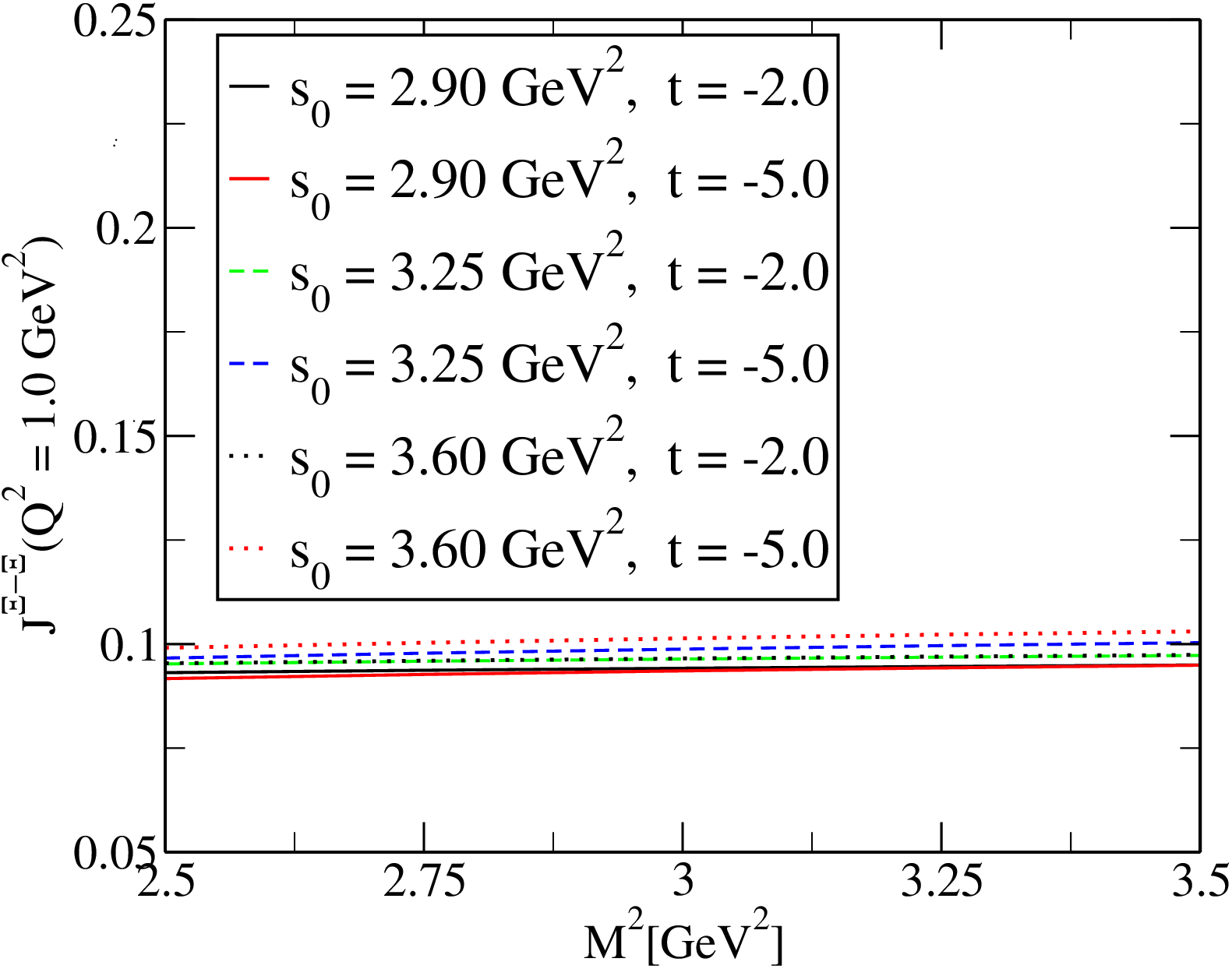}}~~~
\subfloat[]{ \includegraphics[width=0.3\textwidth]{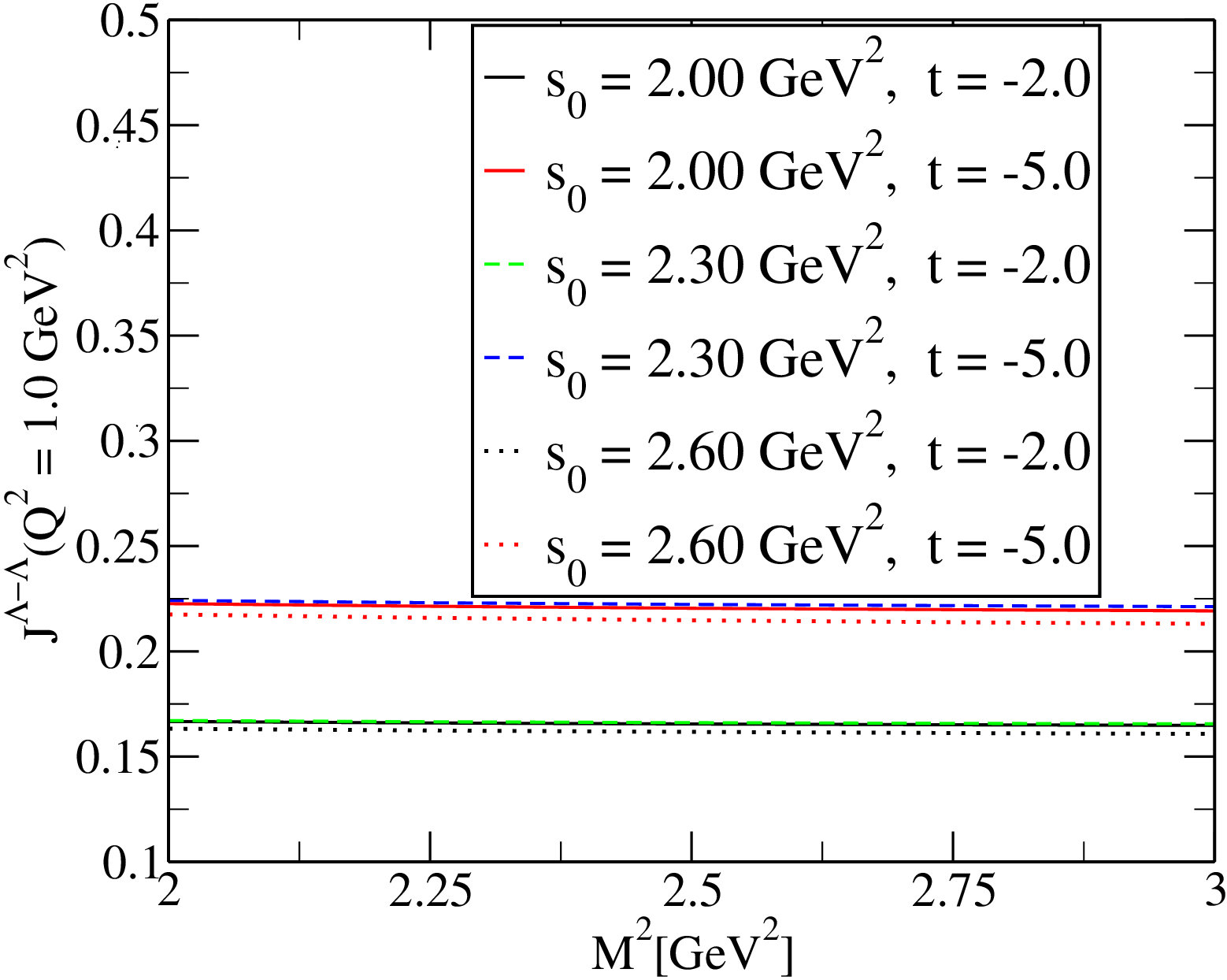}}\\
  \vspace{0.2cm}
   \subfloat[]{ \includegraphics[width=0.3\textwidth]{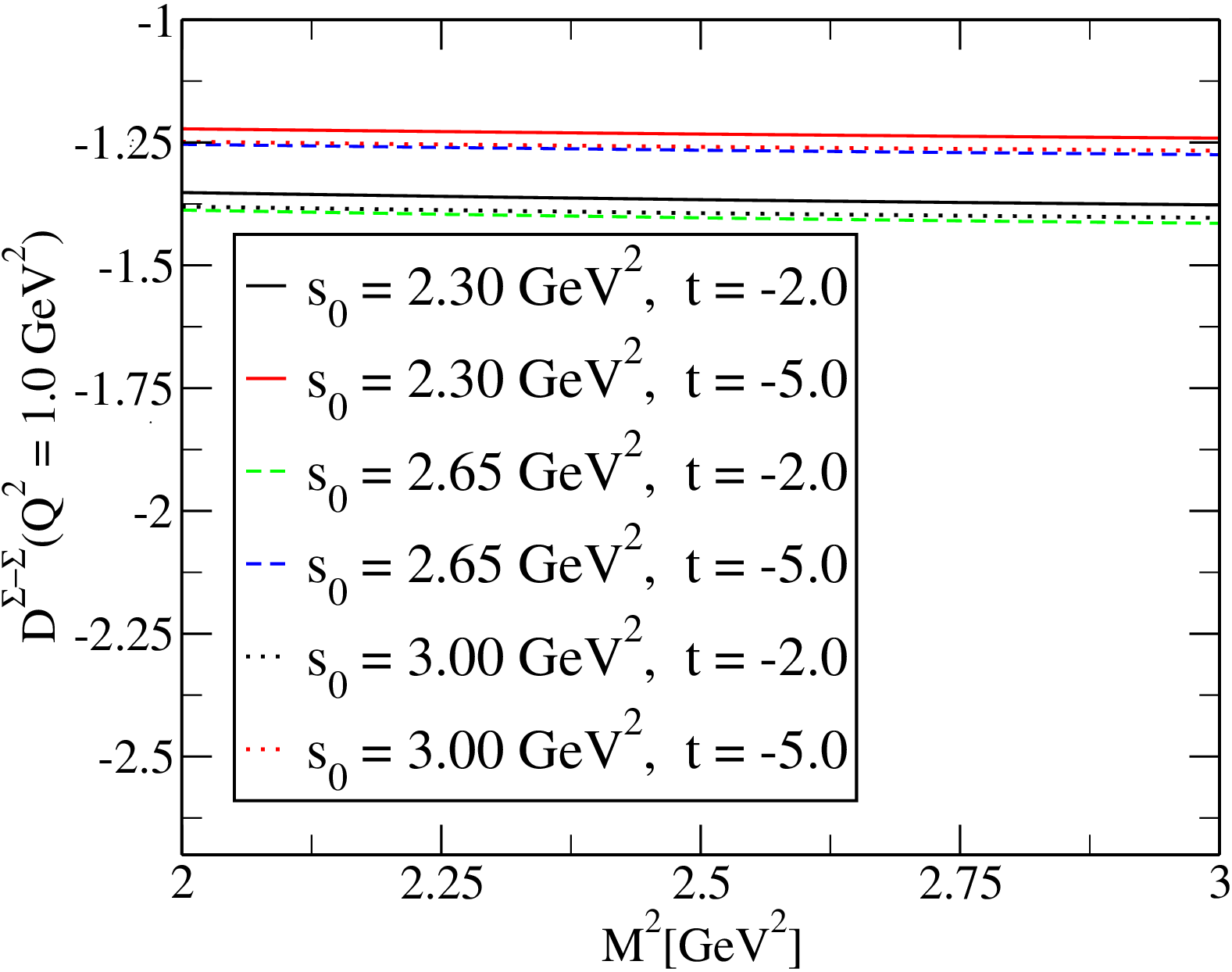}}~~~
   \subfloat[]{ \includegraphics[width=0.3\textwidth]{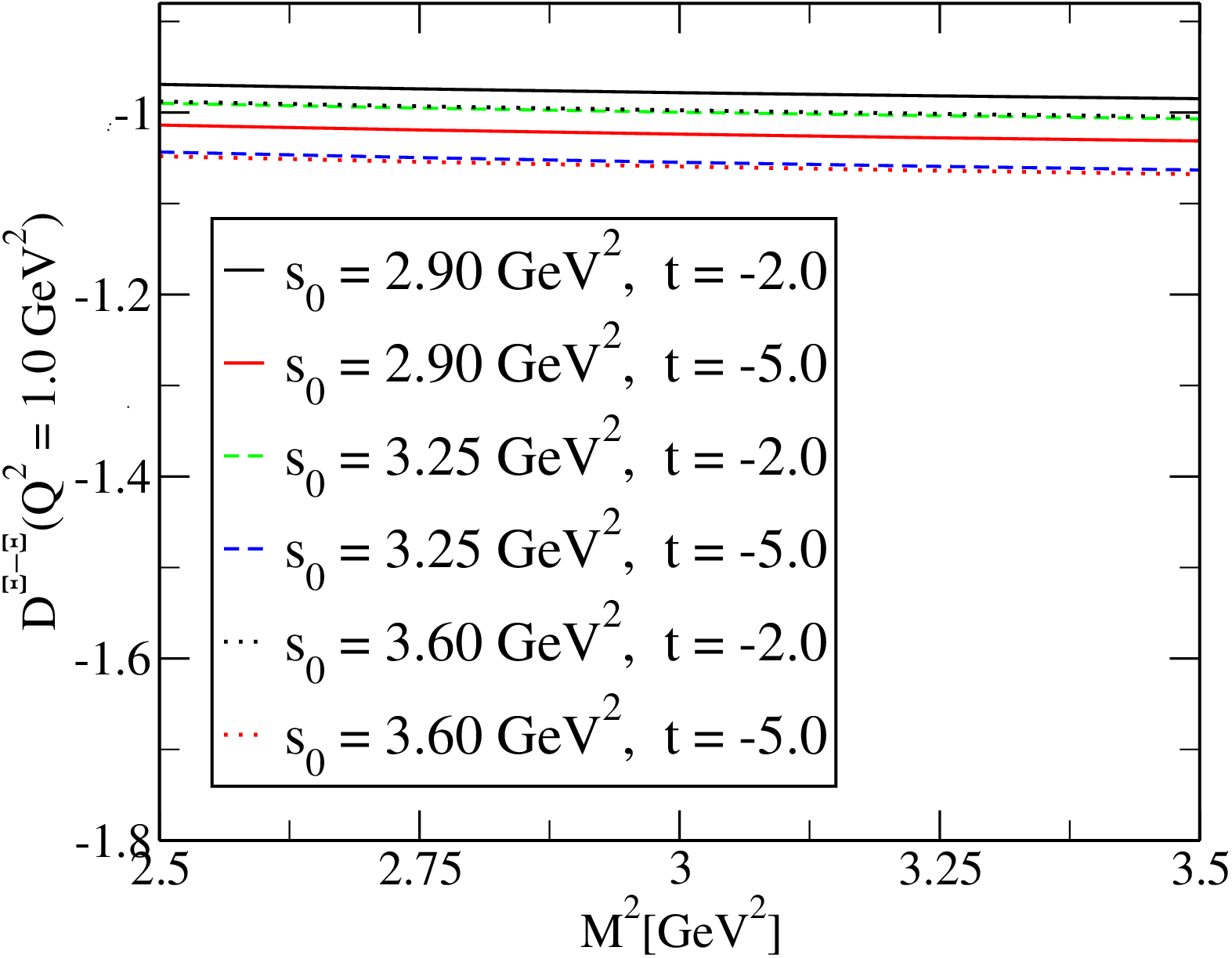}}~~~
   \subfloat[]{ \includegraphics[width=0.3\textwidth]{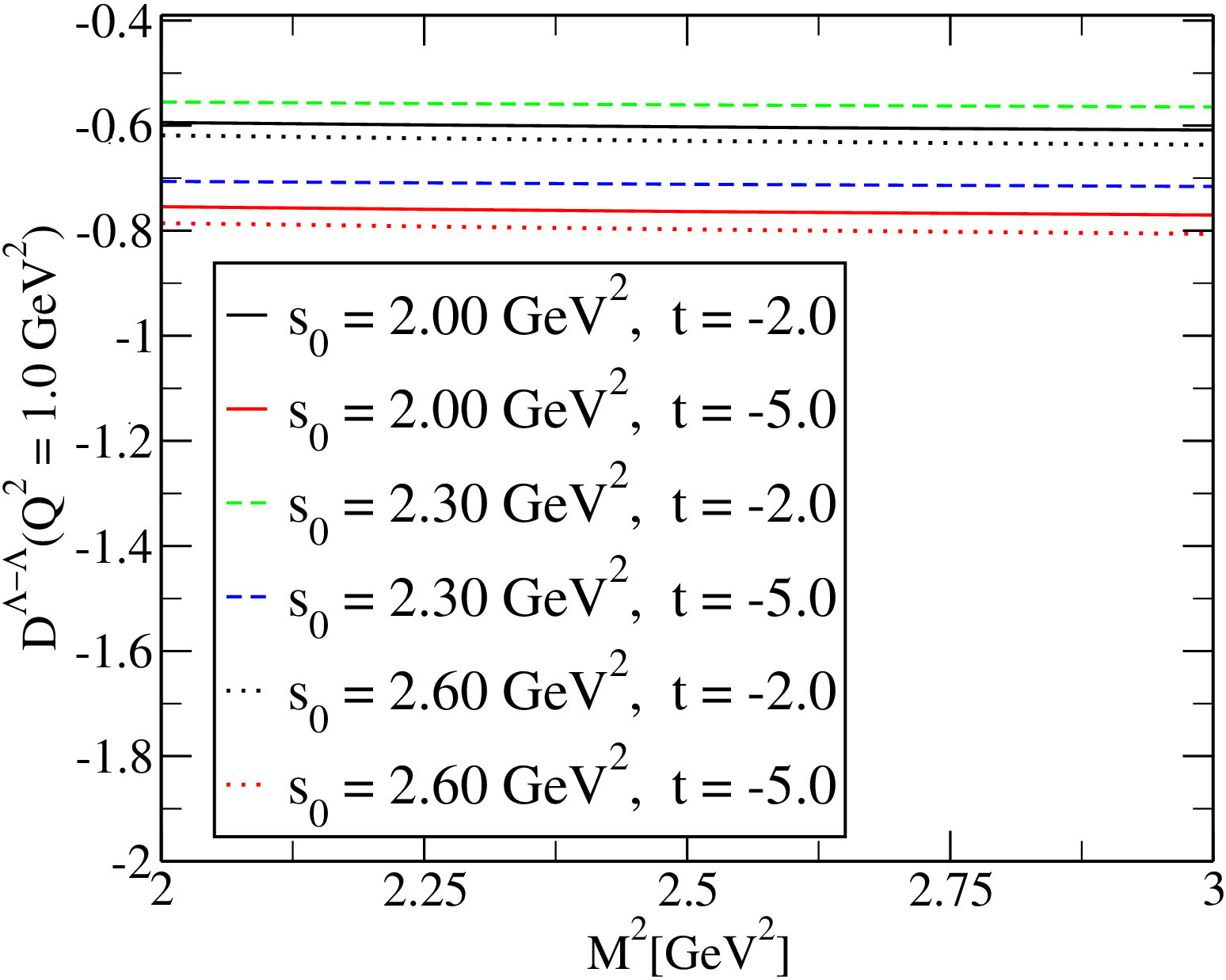}}\\
   \subfloat[]{ \includegraphics[width=0.3\textwidth]{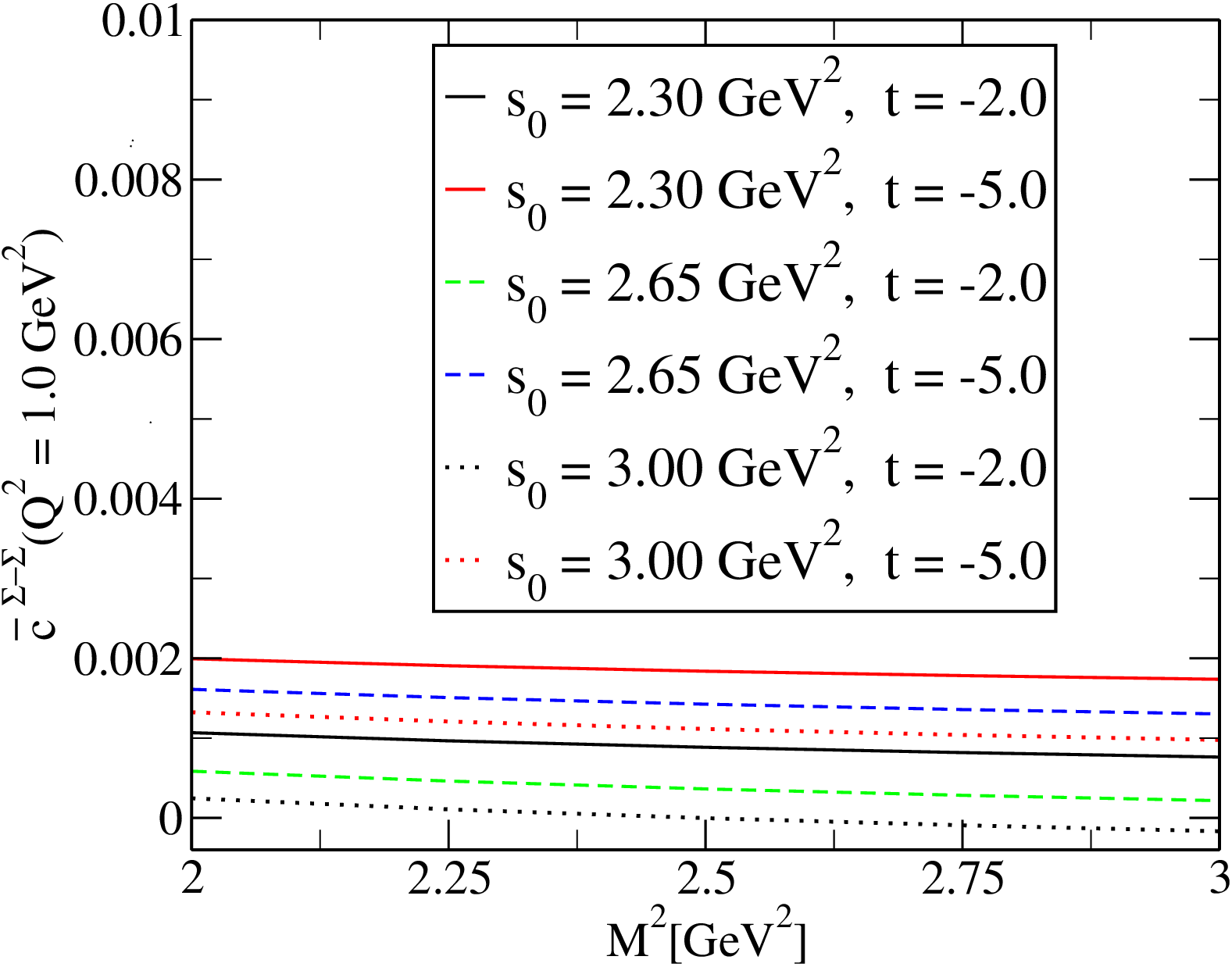}}~~~
 \subfloat[]{ \includegraphics[width=0.3\textwidth]{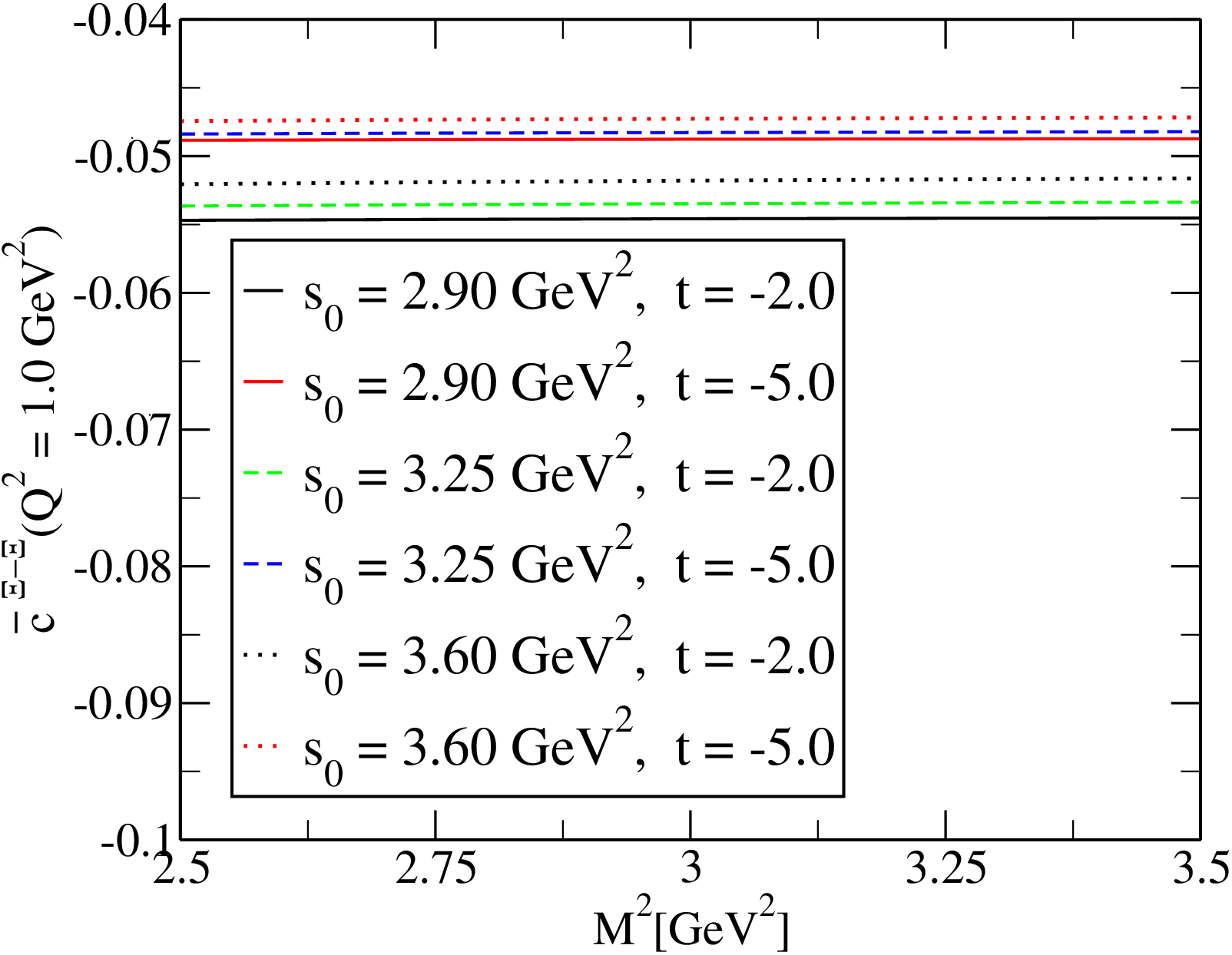}}~~~
 \subfloat[]{ \includegraphics[width=0.3\textwidth]{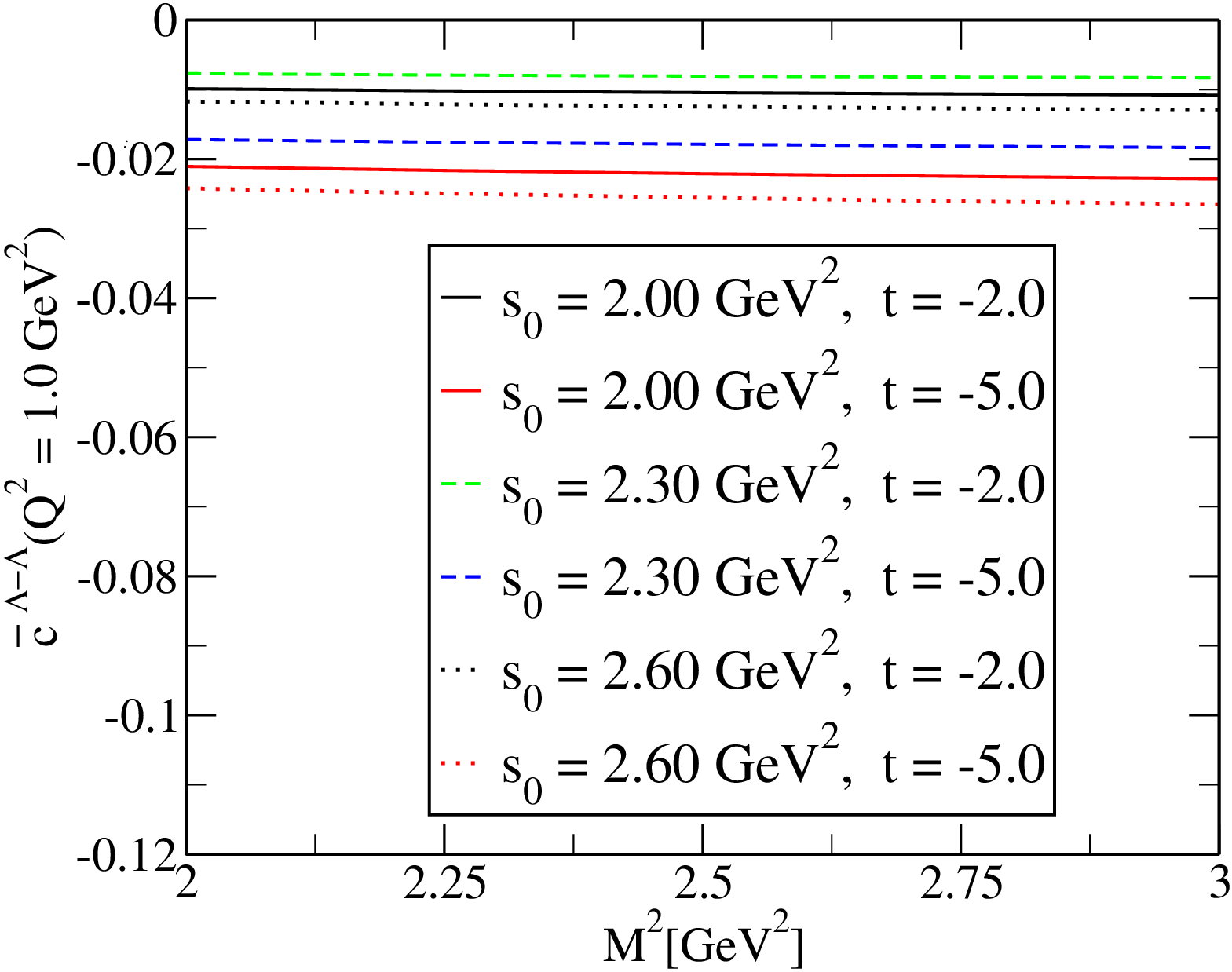}}
 \caption{The dependence of the gravitational form factors of hyperons on $M^2$ at $Q^2$ = 1.0~GeV$^2$ and different values of $ s_0 $ and $ t $ at their working windows.  (a), (d), (g) and (j) for $\Sigma$  baryon,
(b), (e), (h) and (k) for $\Xi$  baryon and; 
(c), (f), (i) and (l) for $\Lambda$  baryon.}
 \label{hyperonMsqfigs}
  \end{figure}
  

 \begin{figure}[t]
\centering
  \subfloat[]{ \includegraphics[width=0.3\textwidth]{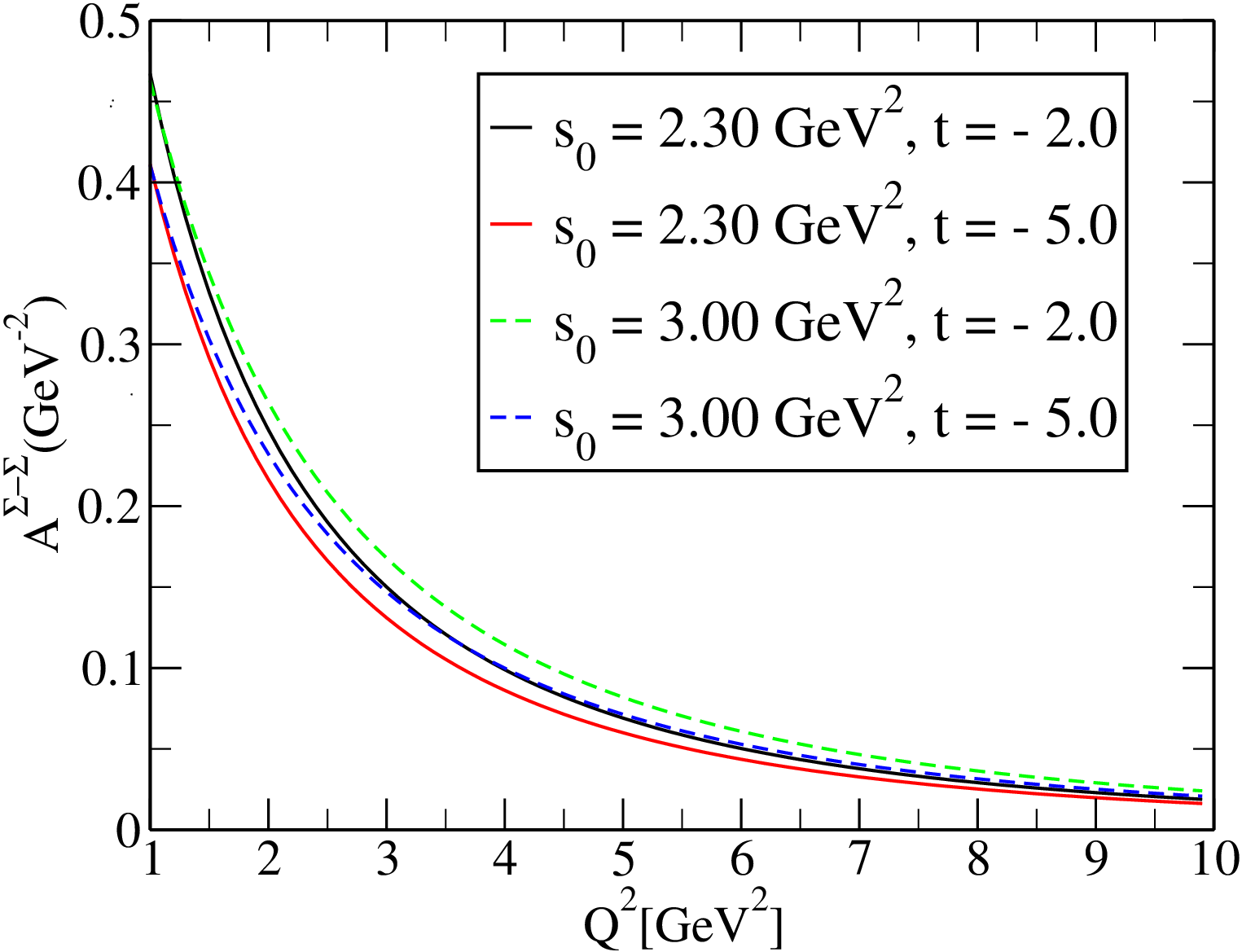}}~~~
 \subfloat[]{ \includegraphics[width=0.3\textwidth]{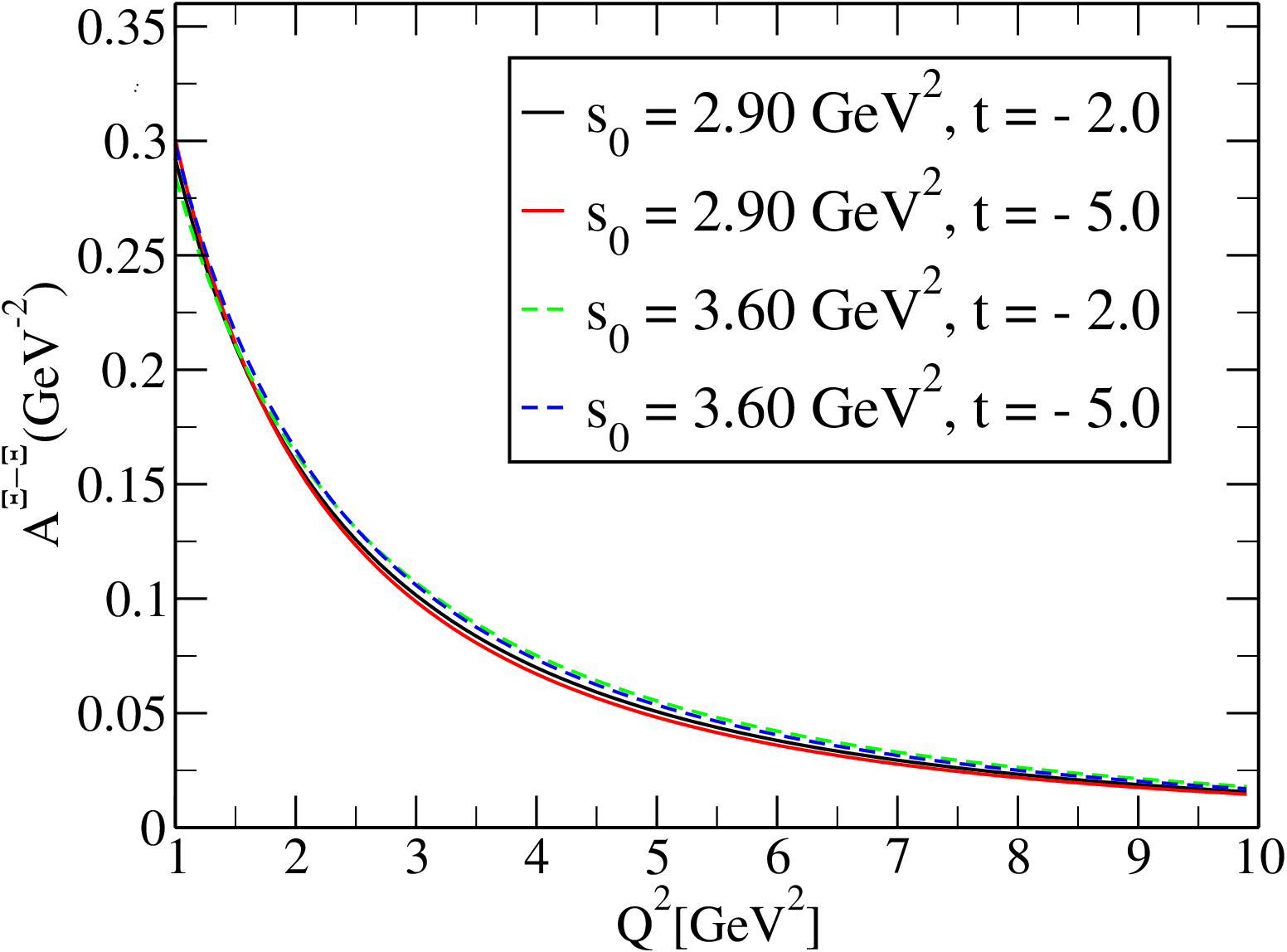}}~~~
 \subfloat[]{ \includegraphics[width=0.3\textwidth]{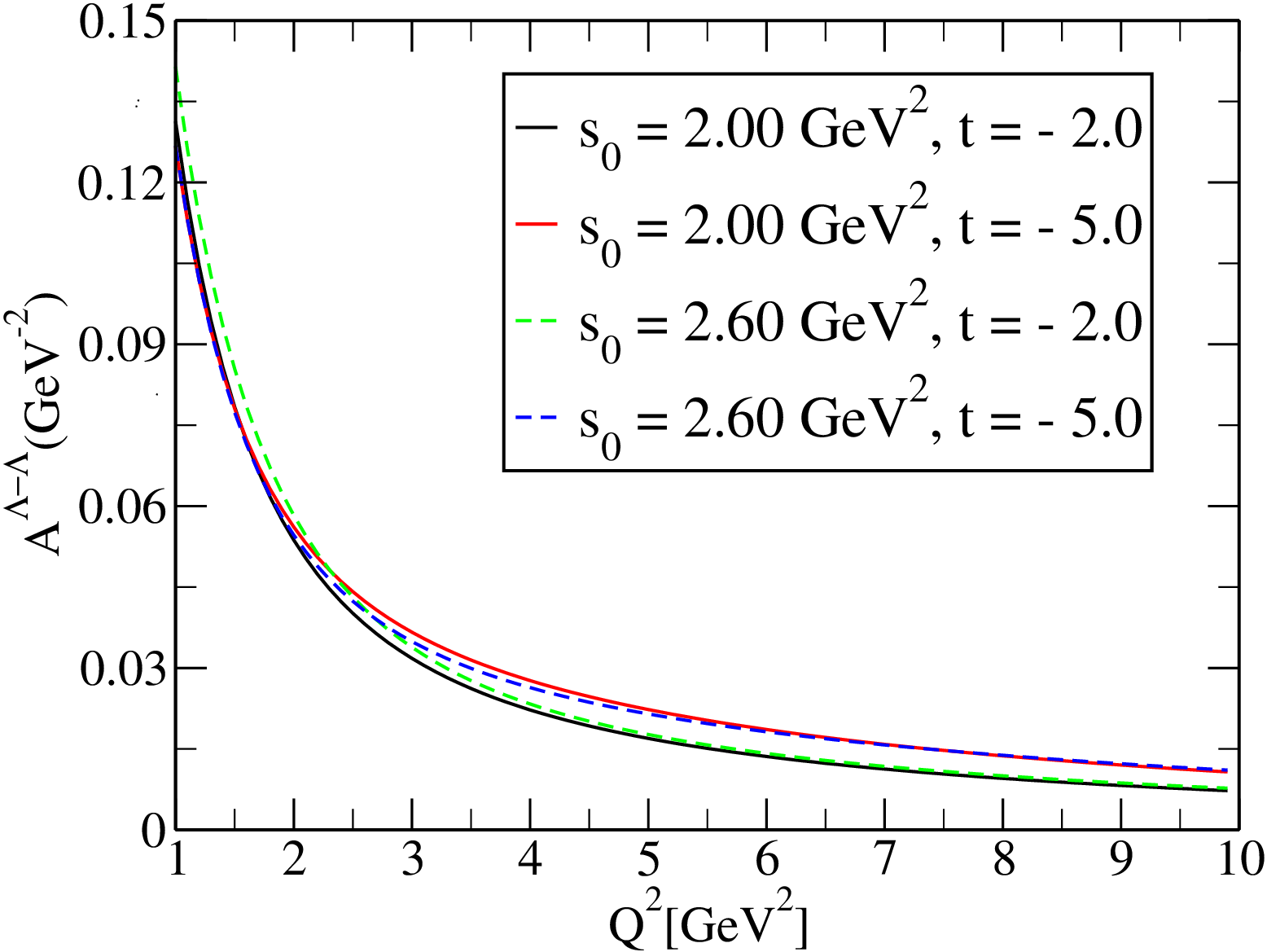}}~~~\\
 \subfloat[]{ \includegraphics[width=0.3\textwidth]{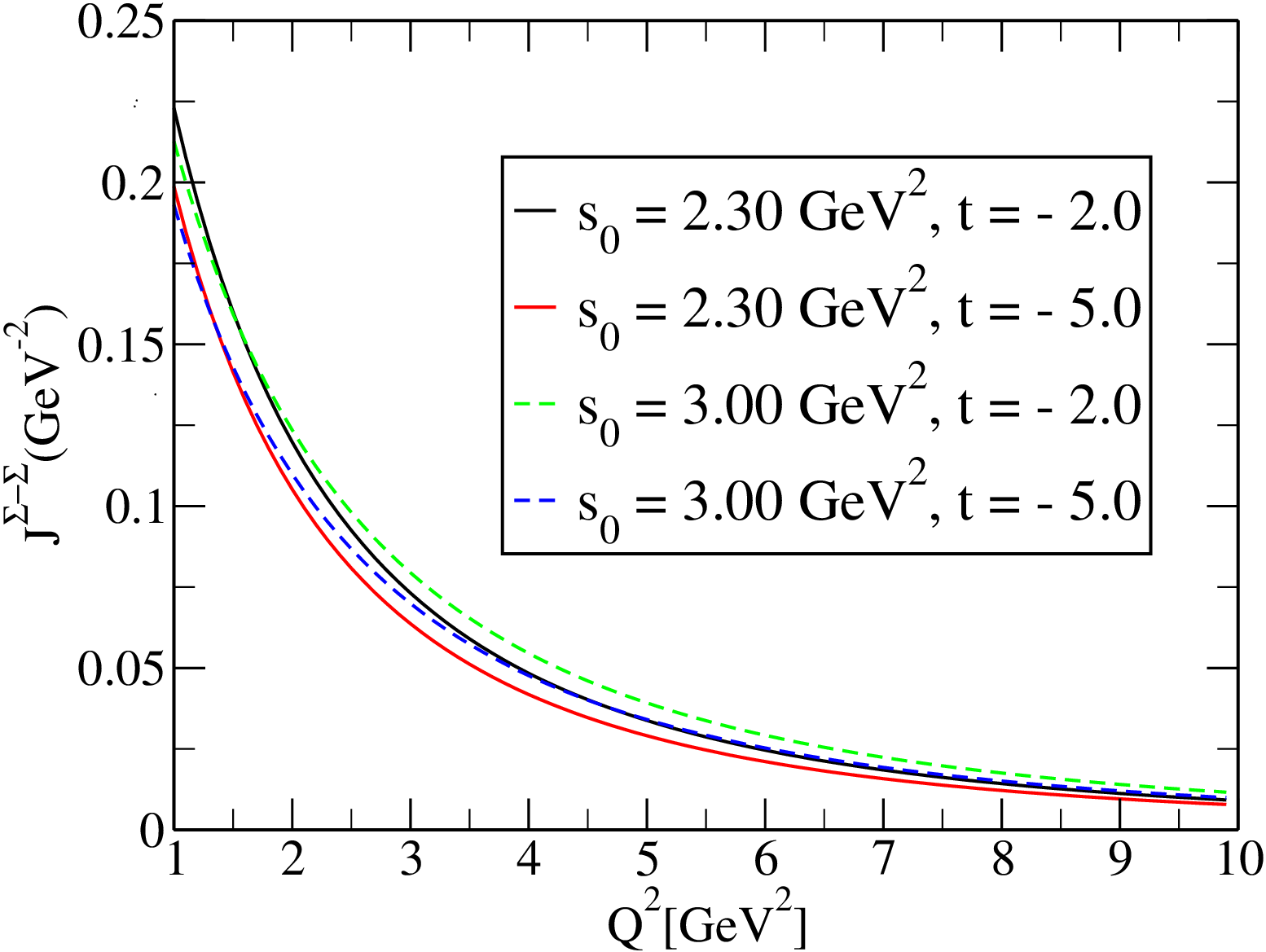}}~~~
\subfloat[]{ \includegraphics[width=0.3\textwidth]{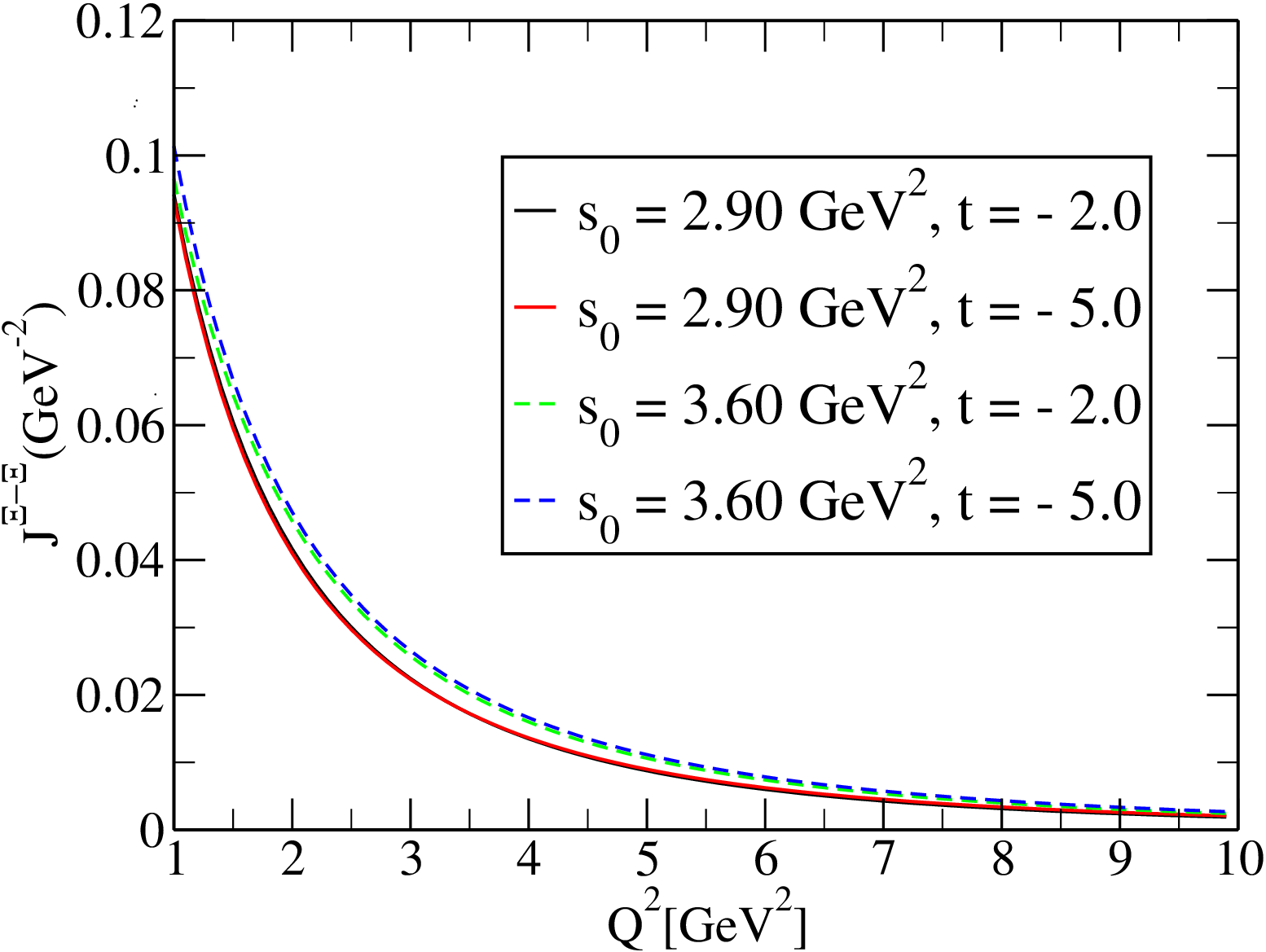}}~~~
\subfloat[]{ \includegraphics[width=0.3\textwidth]{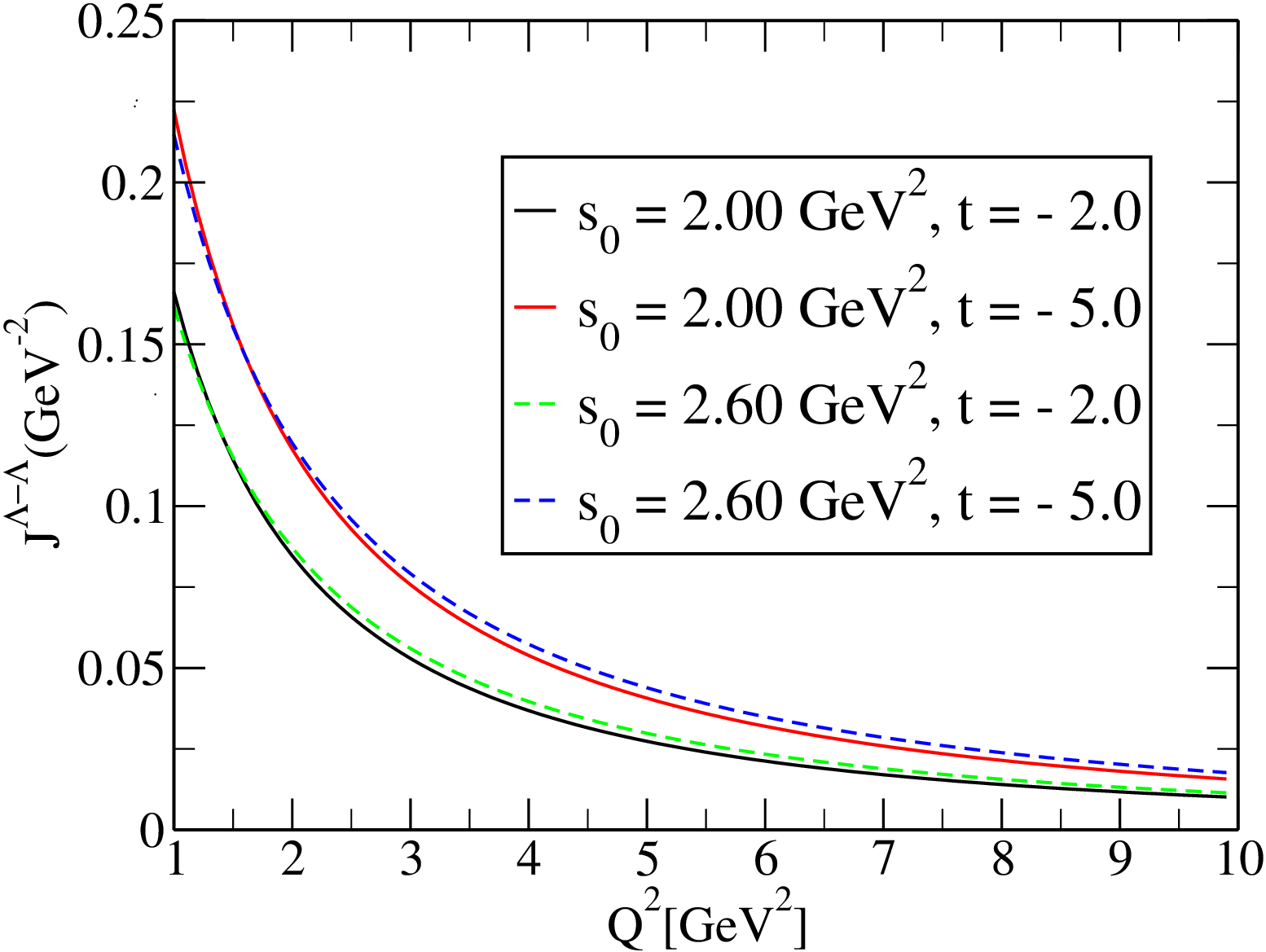}}\\
  \vspace{0.2cm}
   \subfloat[]{ \includegraphics[width=0.3\textwidth]{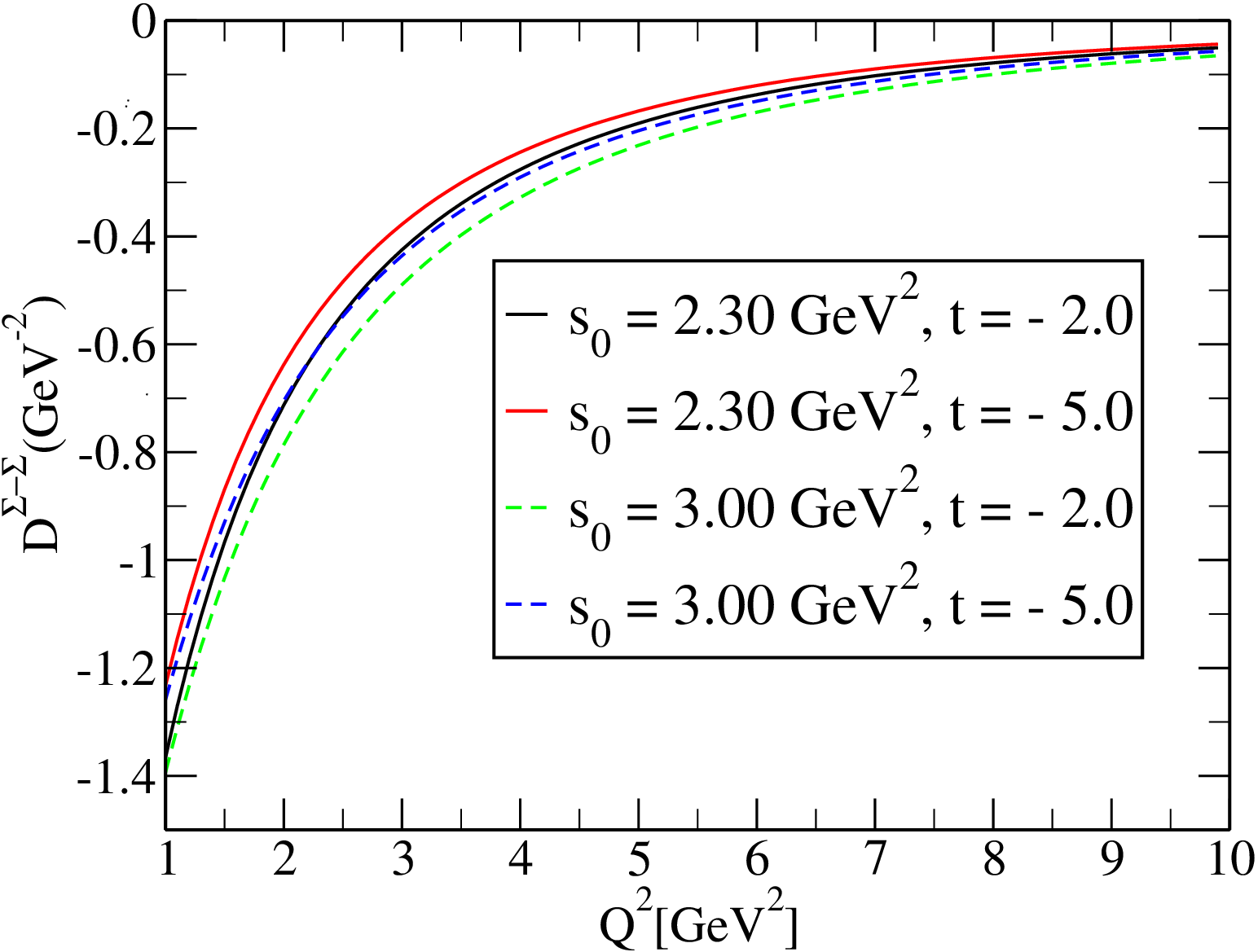}}~~~
   \subfloat[]{ \includegraphics[width=0.3\textwidth]{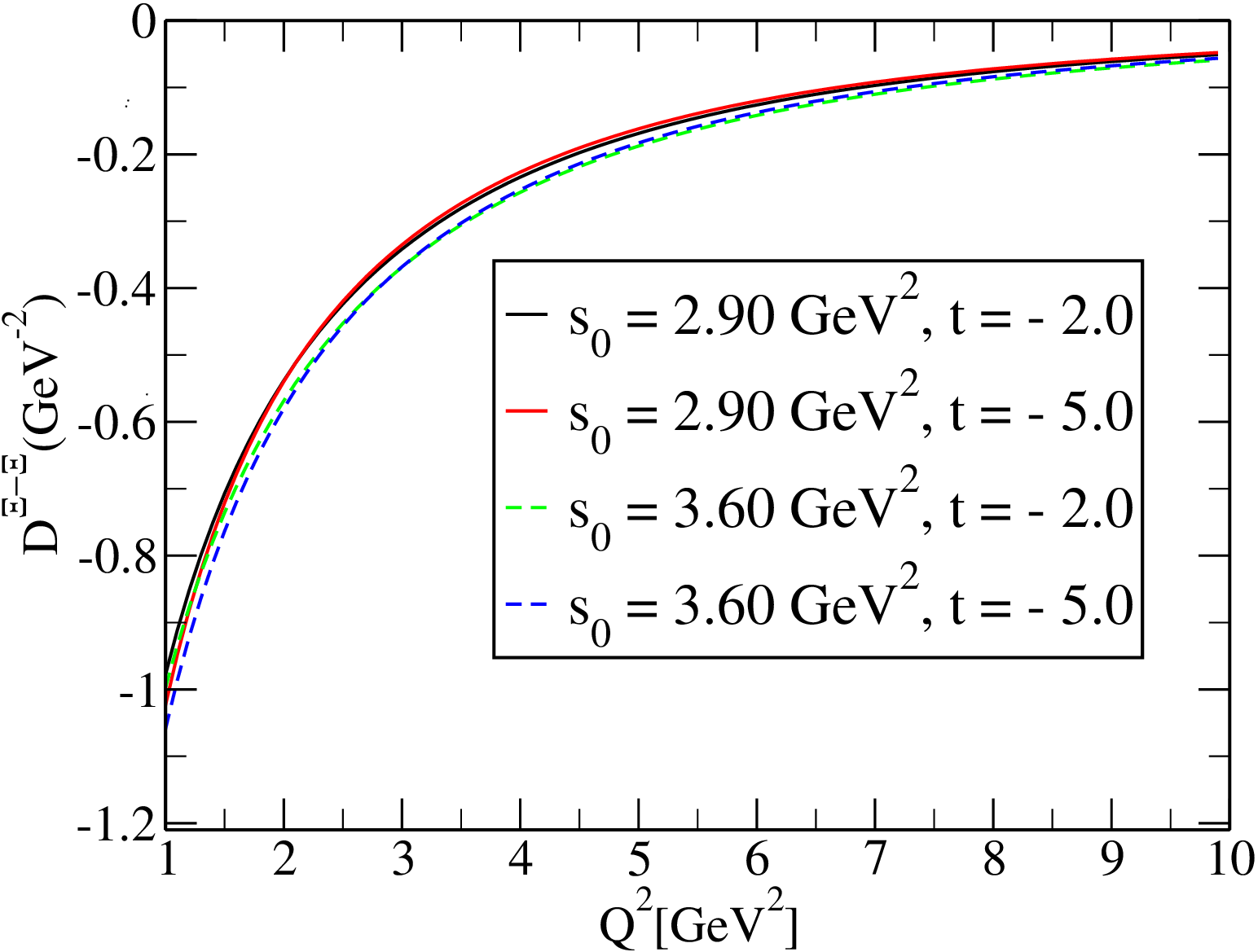}}~~~
   \subfloat[]{ \includegraphics[width=0.3\textwidth]{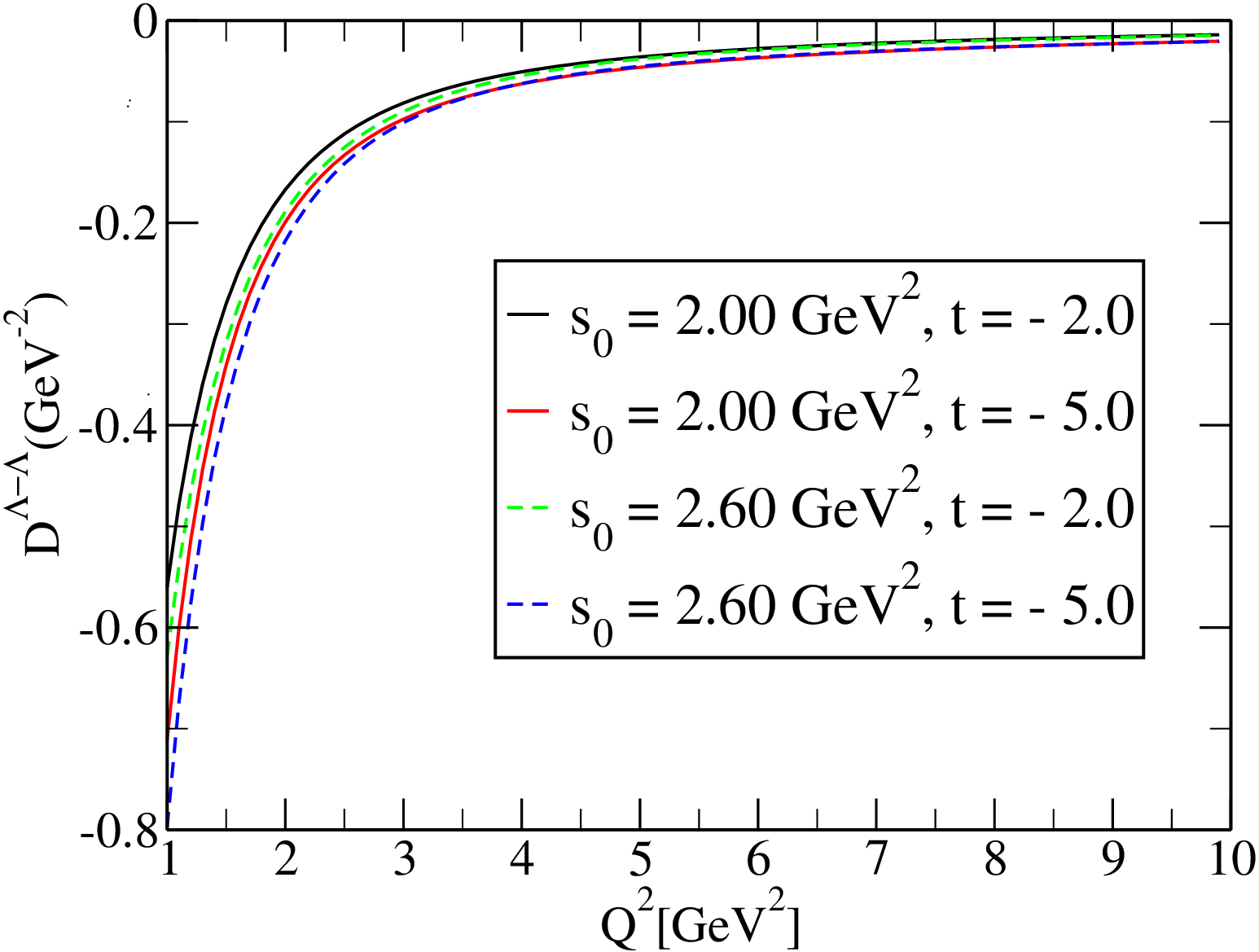}}\\
   \subfloat[]{ \includegraphics[width=0.3\textwidth]{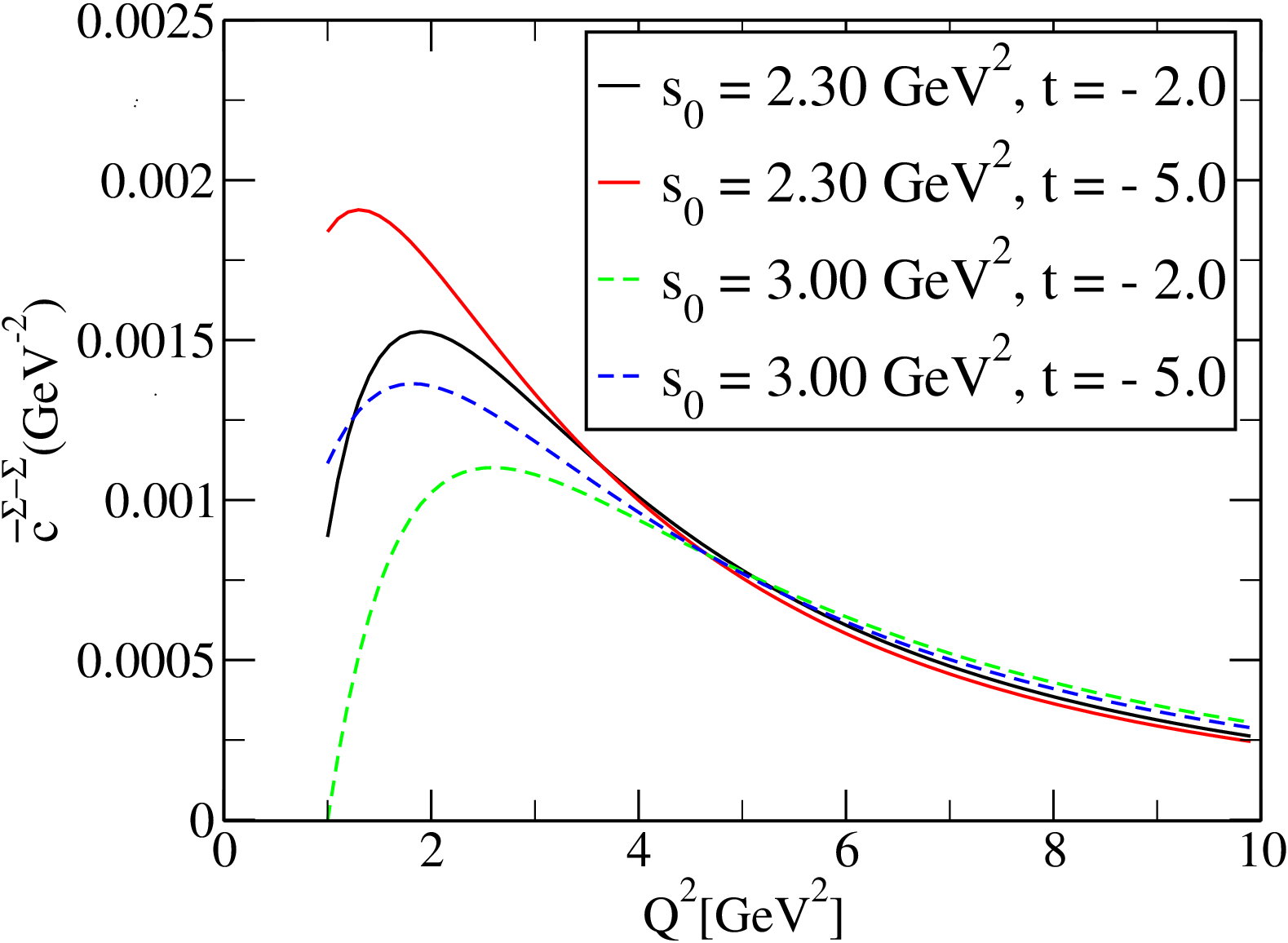}}~~~
 \subfloat[]{ \includegraphics[width=0.3\textwidth]{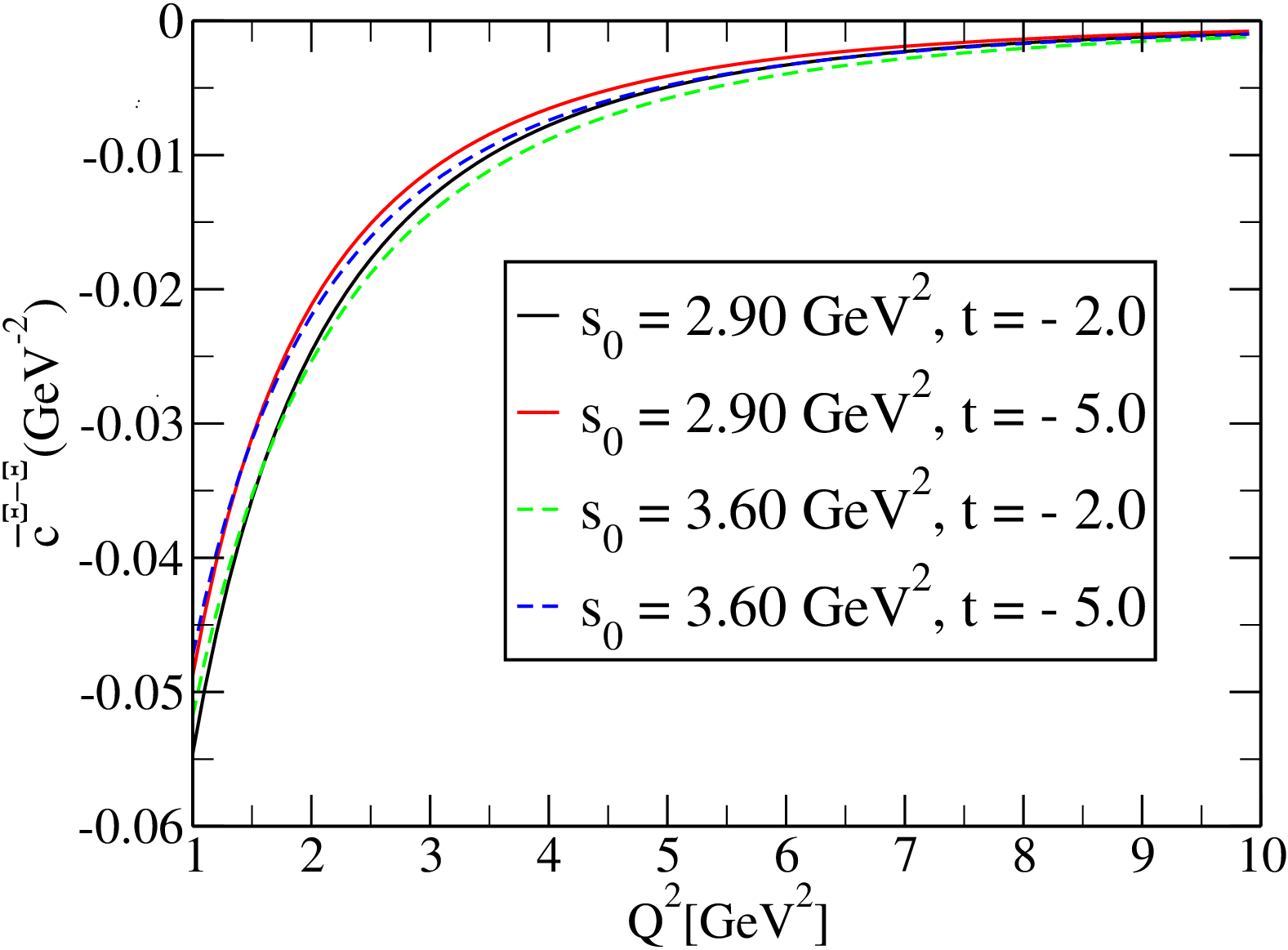}}~~~
 \subfloat[]{ \includegraphics[width=0.3\textwidth]{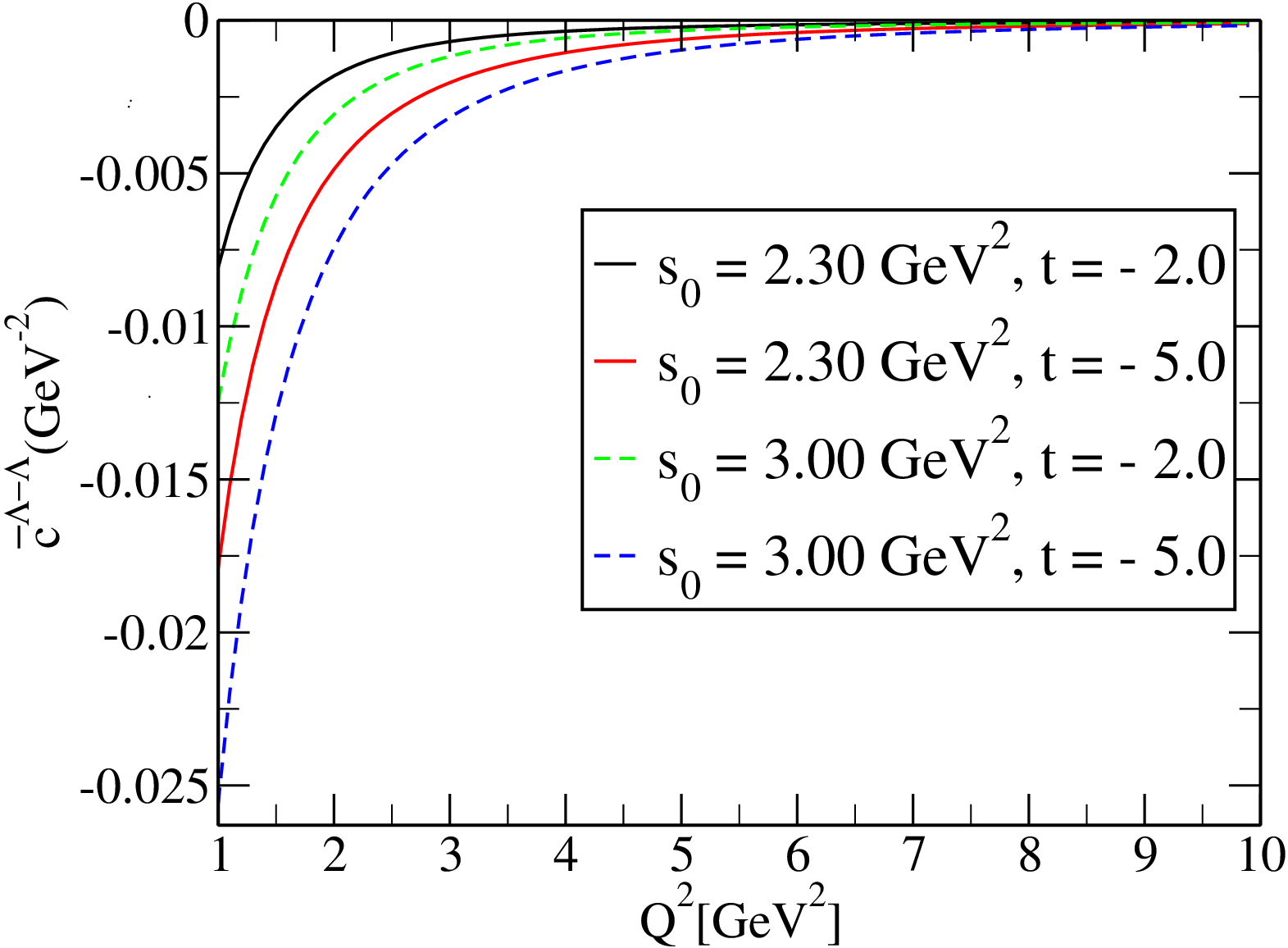}}
 \caption{The dependence of the gravitational form factors of hyperons on $Q^2$ at $M^2$ = 2.5~GeV$^2$ and different values of $ s_0 $ and $ t $ at their working windows. (a), (d), (g) and (j) for $\Sigma$  baryon,
(b), (e), (h) and (k) for $\Xi$  baryon and; 
(c), (f), (i) and (l) for $\Lambda$  baryon.
}
 \label{hyperonQsqfigs}
  \end{figure}

  \end{widetext}

\end{document}